\documentclass[12pt,a4paper,onecolumn,draftcls]{IEEEtran}
\usepackage{epsfig,graphicx,subfigure,psfrag,amsmath,cases}
\usepackage{latexsym,amssymb,amsmath,epsfig,subfigure,algorithm,mathtools}
\usepackage{algorithmic}
\usepackage{color}
\usepackage{url}
\usepackage{scrtime}
\usepackage{epstopdf}
\usepackage{hyperref}
\usepackage{subfigure}
\usepackage{bbding}
\usepackage{multicol}
\usepackage{bm}
\usepackage{bbm}
\usepackage{yhmath}
\usepackage{cite}

\newtheorem{remark}{Remark}

\author{{Ruide Li}, {Zhiqiang Wei}, {Lei Yang}, {Derrick Wing Kwan Ng}, \\
{Jinhong Yuan}, and {Jianping An}
\thanks{R. Li and J. An are with the School of Information and Electronics, Beijing Institute of Technology, China;
Z. Wei, D. W. K. Ng, and J. Yuan are with the School of Electrical Engineering and Telecommunications, the University of New South Wales, Australia;
L. Yang is with Technology and Engineering Center for Space Utilization, Chinese Academy of Sciences.
(e-mail: taiyuanlaide@163.com, zhiqiang.wei@unsw.edu.au, yang.lei@csu.ac.cn, \{w.k.ng, j.yuan\}@unsw.edu.au, an@bit.edu.cn).
This paper was presented in part at IEEE Globecom 2018 \cite{li}\vspace*{-4mm}.
}}

\title{\vspace*{-10mm}Resource Allocation for Secure Multi-UAV Communication Systems with Multi-Eavesdropper}

\newtheorem{Lem}{Lemma}

\newtheorem{T-Prob}{Transformed Problem}

\newtheorem{Proof}{Proof}

\DeclareMathOperator{\maxo}{maximize}
\DeclareMathOperator{\mino}{minimize}

\begin{document}
\maketitle\vspace*{-17mm}
\begin{abstract}
  \vspace*{-4mm}
  In this paper, we study the resource allocation and trajectory design for secure unmanned aerial vehicle (UAV)-enabled communication systems, where multiple multi-purpose UAV base stations are dispatched to provide secure communications to multiple legitimate ground users (GUs) in the existence of multiple eavesdroppers (Eves).
  Specifically, by leveraging orthogonal frequency division multiple access (OFDMA), active UAV base stations can communicate to their desired ground users via the assigned subcarriers while idle UAV base stations can serve as jammer simultaneously for communication security provisioning.
  To achieve fairness in secure communication, we maximize the average minimum secrecy rate per user by jointly optimizing the communication/jamming subcarrier allocation policy and the trajectory of UAVs, while taking into account the constraints on the minimum safety distance among multiple UAVs, the maximum cruising speed, the initial/final locations, and the existence of cylindrical no-fly zones (NFZs).
  The design is formulated as a mixed integer non-convex optimization problem which is generally intractable.
  Subsequently, a computationally-efficient iterative algorithm is proposed to obtain a suboptimal solution.
  Simulation results illustrate that the performance of the proposed iterative algorithm can significantly improve the average minimum secrecy rate compared to various baseline schemes.\vspace*{-5mm}
\end{abstract}


\section{Introduction}\vspace*{-1.5mm}

The rapid growing demand on wireless communication services, e.g. ultra-high data rates and massive connectivity \cite{j_an}, has fueled the development of wireless networks and the mass productions of wireless devices.
Despite the fruitful research in the literature for providing ubiquitous services, the performance of wireless systems is limited by the users with poor channel conditions  \cite{K_Yang,X_Gao}.
Fortunately, owing to the high flexibility and low cost in deployment of unmanned aerial vehicles (UAVs), UAV-enabled communication offers a promising solution to tackle these challenges \cite{wong_2017_Key_Technologies_for_5G}.
In particular, the high mobility of UAVs facilitates the establishment of strong line-of-sight (LoS) links to ground users (GUs).
Hence, in recent years, numerous applications of UAV-enabled communication have emerged dramatically not only in the military domain, but also in the civilian and commercial domains, such as disaster relief, archeology, pollution monitoring, commodity delivery, etc. \cite{Zeng_2016_Unmanned_Aerial_Vehicles_Magazine}.
Besides, several world-leading industrial companies, such as Facebook, Google, and Qualcomm, have made advancements on their journey to deliver high-speed internet from the air by UAVs \cite{li}.
As a result, the investigation of deploying UAVs for assisting wireless networks has recently received significant attention from the academia, such as mobile relays \cite{Zeng_2016_Throughput_Maximization_for_UAV,xiao2018secrecy}, aerial mobile base stations \cite{Device_to_Device_Communications_Mozaffari,Robust_sun_2019}, and UAV-enabled information dissemination and data collection \cite{li,Zeng_2017_Completion_Time_Minimization_in_UAV}.

In practical systems, although the nature of strong LoS link grants UAV-based communication as an appealing approach to provide ubiquitous high-data rate wireless service, it also makes the communication between a UAV and ground users more susceptible to be intercepted by potential eavesdroppers (Eves) \cite{Sun_2019_Physical}.
Therefore, it imposes various fundamental challenges for secure UAV communication provisioning \cite{wu2019safeguarding}.
To meet this emerging need, secure UAV communication systems with a single-UAV was studied in \cite{xiao2018secrecy,Zhang2018Securing,cui2018robust,Yuanxin_Cai} with different system settings.
However, due to the stringent requirements on UAV's size, weight, and power (SWAP), the  performance achieved by deploying a single-UAV is still limited \cite{Zeng_2016_Unmanned_Aerial_Vehicles_Magazine}.
To achieve a higher efficiency in secure communications, multi-UAV cooperation was adopted in \cite{a_li_cooperative_2019,Zhong_Cooperative_Jamming,Y_Li_Access,Lee_cooperative_jamming}.
In particular, a jammer UAV can fly close to a potential eavesdropper based on demand by leveraging its mobility and opportunistically transmits artificial noise signal deliberately to combat the eavesdropping channels \cite{a_li_cooperative_2019}.
To improve the system security performance, \cite{Zhong_Cooperative_Jamming} and \cite{Y_Li_Access} presented a cooperative jamming approach to safeguard the UAV's communication by exploiting artificial jamming transmission from other friend UAVs in the existence of a single-eavesdropper.
With the consideration of fairness in two-UAV secure communications, \cite{Lee_cooperative_jamming} investigated the joint power allocation and trajectory design for the maximization of the minimum secrecy rate per user when one UAV is dispatched to convey confidential messages to a ground user where another cooperative UAV transmits a jamming signal.
However, in \cite{a_li_cooperative_2019,Y_Li_Access,Zhong_Cooperative_Jamming,Lee_cooperative_jamming}, the role of the UAV is fixed where a communication/jamming UAV can only provide either communication/jamming signal during the whole time horizon.
In contrast, a multi-purpose UAV, which can dynamically serve as a communication UAV or a jamming UAV, provides a high flexibility in trajectory design for secure UAV communications.
For instance, when a communication UAV flies closer to an eavesdropper,
it can switch its role from a communication UAV to a jamming UAV for improving the system performance of secure communication.
However, an efficient algorithm for optimizing resource allocation and the trajectory of multi-purpose UAVs has not been reported, yet.
Moreover, \cite{xiao2018secrecy,Zhang2018Securing,a_li_cooperative_2019,Zhong_Cooperative_Jamming} only considered the scenario of a single-user and one eavesdropper.
On the other hand, although a single-user with the existence of multiple eavesdroppers in UAV-enabled communication systems was investigated in \cite{cui2018robust,Yuanxin_Cai,Y_Li_Access,Lee_cooperative_jamming}, these results are not applicable to most of practical and important scenarios in the present of multiple desired users.
To the best of our knowledge, secure communication problem in a more general scenario, the coexistence of multiple users and multiple eavesdroppers, is very challenging and  still remained to be explored.

To unleash the potential performance of UAV-enabled communications, trajectory design or path planning has been a major research area in the existing literature.
For example, \cite{Zeng_2016_Throughput_Maximization_for_UAV} optimized the trajectory of a UAV to maximize the system throughput of a single-user while taking into account its maximum mobility.
Authors in \cite{Zhang_2017_Securing_UAV_Communications} investigated the UAV's trajectory design to guarantee secure air-to-ground communications.
Although the UAV trajectory has been designed with different practical considerations \cite{cui2018robust,Yuanxin_Cai,Y_Li_Access,Lee_cooperative_jamming,a_li_cooperative_2019,Zhong_Cooperative_Jamming}, e.g. UAV's velocity, initial/final locations, and energy efficiency, physical geometric restriction is remained to be investigated in UAV trajectory design.
For example, due to regulations for military, security, safety or privacy reasons, there are some no-fly zones (NFZs) where the flight of UAVs over those regions is prohibited \cite{Valavanis:2014:HUA:2692452,zhao_chen_yu_2017,Gao_Access_NFZ}.
Therefore, the authors of \cite{Valavanis:2014:HUA:2692452} proposed a control mechanism based on the geometrical tangential method of control theory to avoid a UAV flying into the NFZ.
However, their proposed method only focused on the cruise constraint of the UAV due to NFZ which did not take into account any security concerns of air-to-ground data communication.
Thus, with the consideration of NFZs, in our previous works \cite{li}, \cite{Multiuser_MISO_NFZ}, we investigated the designs of resource allocation algorithm for UAV-enabled communication systems, where a UAV is dispatched to provide communications to multiple ground users in the present of multiple NFZs.
Also, with the consideration of NFZs, the authors in \cite{Gao_Access_NFZ} studied the UAV secure communication system design to serve a single-legitimate user with the presence of a malicious eavesdropper.
However, the existence of NFZs complicates the design of resource allocation and the results from \cite{li,Valavanis:2014:HUA:2692452,zhao_chen_yu_2017,Gao_Access_NFZ,Multiuser_MISO_NFZ} are not applicable to role switching design among UAVs.

Based on the aforementioned observations, in this paper, we consider a multi-UAV-enabled orthogonal frequency division multiple access (OFDMA) communication systems, where multiple rotary-wing UAVs base stations are dispatched to provide communications to multiple ground users with the existence of multi-eavesdropper and NFZs.
We jointly design the resource allocation, trajectory design, and role selection for secure communication.
In particular, the role of each UAV can be switched dynamically between serving as a jamming UAV or information UAV in each time slot and subcarrier.
The design optimization problem is non-convex and generally intractable.
To handle the above challenges, we first transform the original problem into its equivalent problem, which facilitates the application of alternating optimization for obtaining a suboptimal solution.
In particular, the original optimization problem is divided into three subproblems, i.e., communication resource allocation, jamming policy, and UAVs' trajectories design, which to be solved iteratively.
In each iteration, the communication resource allocation is designed by solving its Lagrangian dual problem.
To derive jamming policy and UAVs' trajectories, a suboptimal iterative algorithm is proposed by utilizing successive convex approximation (SCA) techniques\cite{Y_Li_Access,Zhong_Cooperative_Jamming,Lee_cooperative_jamming} with a fast convergence.

The remainder of this paper is organized as follows.
Section II introduces the system model and the problem formulation for the considered cooperative multi-UAV enabled wireless system.
In Section III, we propose an efficient iterative algorithm based on the Lagrange dual problem and SCA techniques which can obtain a suboptimal solution of the design problem at hand.
Section V provides numerical results to demonstrate the performance of the proposed algorithms.
Finally, the paper is concluded in Section VI.

Notations:
$\|\cdot\|$ denotes the vector norm.
$[x]^+ = \max \{0,x\}$. $[\cdot]^{\mathrm{T}}$ denotes the transpose operation.
For a vector $\mathbf{a}$, $\|\mathbf{a}\|$ represents its Euclidean norm.

\begin{figure}[t]
  \centering
  \includegraphics[width=3.3 in]{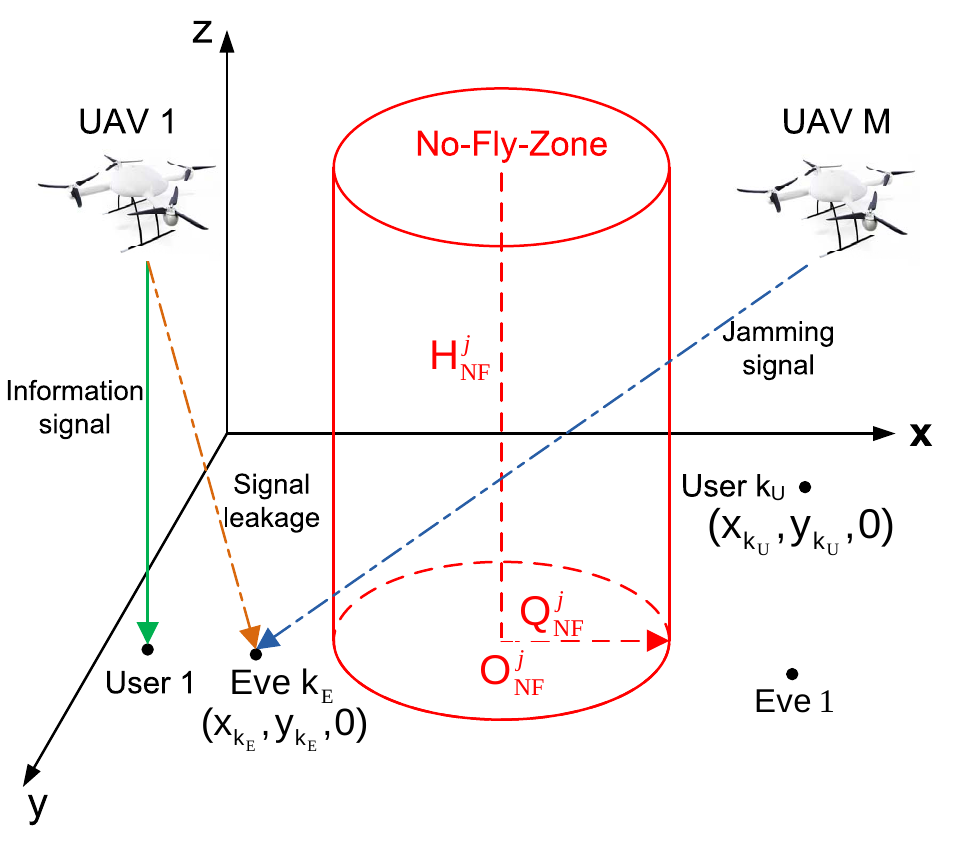}\vspace*{-8mm}
  \caption{Secure multi-UAV communication systems with the existence of multi-eavesdropper and a NFZ.}\vspace*{-10mm}
  \label{fig:UAV communication frame}
\end{figure}

\vspace*{-2mm}
\section{System Model}\vspace*{-1.5mm}

We consider a wireless communication system consisting of $K_U$ downlink users, $K_E$ eavesdroppers, $M$ rotary-wing UAVs denoted by set $k_U \in \mathcal{K}_U \triangleq \{1, \cdots, K_U\} $, $k_E \in \mathcal{K}_{E} \triangleq \{1, \cdots, K_E\} $, and $m\in\mathcal{M} \triangleq \{1, \cdots, M\} $, respectively.
{\color{black}
In the system, we adopt the OFDMA scheme for serving multiple downlink users with different subcarriers at the same time \cite{li,Wu_OFDM_UAV_2017}.
}
Besides, the system bandwidth is divided equally into $N_\mathrm{F}$ orthogonal subcarriers, which is denoted by set $i \in \mathcal{N}_\mathrm{F} \triangleq \{1, \cdots, N_\mathrm{F}\} $.
As there are multiple eavesdroppers in the system, we exploit the geometrically distributed nature of UAVs and advocate the use of their dynamical role switching for guaranteeing secure communication.
To be more specific, on each subcarrier, at a given time if a UAV\footnote{We assume that there are secure backhaul links for conveying a user's data between UAV and a core network which can be established by e.g. out-of-band links \cite{Device_to_Device_Communications_Mozaffari,Robust_sun_2019}.} is selected for information transmission, the other UAVs can serve as jamming UAVs for protecting the communication via artificial noise transmission, as shown in Fig. \ref{fig:UAV communication frame}.
The policy adapts the communication and jamming roles of UAVs which enables a highly flexible resource allocation for improving the system performance{\color{black}\footnote{{\color{black}The cost of the proposed role switching is mainly on the required signaling overhead for the coordination between the cooperative UAVs.}}}.

\vspace*{-6mm}
\subsection{Signal Model}\vspace*{-2mm}

We express the locations of all users, eavesdroppers, and UAVs in a three-dimensional (3D) Cartesian coordinate system.
For the ease of design, the total time $T$ is discretized into $N$ time slots with equal-duration, i.e., $\delta t = T/N$, which is small enough such that the distance between the UAV and the ground user within each time slot can be treated as a constant\footnote{The discretized model is commonly adopted in the literature which facilitates the design of resource allocation for UAV-enabled communication systems \cite{cui2018robust,Zhang2018Securing,Zeng_2016_Throughput_Maximization_for_UAV,
xiao2018secrecy,Yuanxin_Cai,Zhong_Cooperative_Jamming,Lee_cooperative_jamming,a_li_cooperative_2019}.}.
Furthermore, we adopt $n$ as the time slot index where $n \in \mathcal{N} \triangleq \{1, \cdots, N\}$.
Thus, the ground projected trajectory of UAV $m$, $\mathbf{q}_m(t)=[x_m(t),y_m(t)]$, $0\le t\le T$, over the time $T$ can be approximated by a sequence $\{\mathbf{q}_m[n] = \left[ x_m[n],y_m[n] \right]^T\}_{n=1}^N$, where $\mathbf{q}_m[n] \triangleq \mathbf{q}_m(n\delta t)$, $\forall n, m$, denotes the horizontal location of UAV $m$ in time slot $n$.
Besides, the maximum speed of each UAV are denoted as $v_{\rm{max}}$ in meters per second (m/s) and the UAV's maximum aviation distance in each time slot is $V=v_{\rm{max}}\delta t$ in meters.
In particular, without lost of generality, we assume that the initial location
and the final location for UAV $m$ projected on the ground is $\mathbf{q}_{m}[0] = [x_m[0],y_m[0]]^T$ and $\mathbf{q}_m[N] = [x_m[N],y_m[N]]^T$, and the horizontal coordinates for user $k_U$ and Eve $k_E$ are denoted by $\mathbf{w}_{k_U} = [x_{k_U},y_{k_U}]^T$ and $\mathbf{w}_{k_E} = [x_{k_E},y_{k_E}]^T$,  respectively.
Then, the distance between UAV $m$ and user $k_U$ in time slot $n$ can be written as\vspace*{-2mm}
\begin{equation}\label{eqn:distance of communication}
  d_{m,k_U}[n] = \sqrt{ || \mathbf{q}_m[n] - \mathbf{w}_{k_U} ||^2 + H^2 },\;  \forall n, m, k_U,\vspace*{-1mm}
\end{equation}
where $H$ in meters is the constant flying altitude of each UAV for satisfying some safety regulations.
Similarly, the distance between UAV $m$ and Eve $k_E$ in time slot $n$ is given by\vspace*{-2mm}
\begin{equation}\label{eqn:distance of communication}
  d_{m,k_E}[n] = \sqrt{ || \mathbf{q}_m[n] - \mathbf{w}_{k_E} ||^2 + H^2 },\;  \forall n, m, k_E.
\end{equation}

\vspace*{-6mm}
\subsection{No-Fly-Zone Model}\vspace*{-1.5mm}

In practice, UAVs flying over some specific regions, known as NFZ, such as airports, prisons, military locations, and etc. are prohibited.
Hence, a practical UAV-based communication system should design the UAV trajectory with the consideration of NFZs.
In this paper, we assume that there are ${N_\mathrm{NF}}$ non-overlapped NFZs.
Specifically, to guarantee the effectiveness of our UAV trajectory design, we define NFZ $j\in\{1,...,N_\mathrm{NF}\}$, as a cylindrical volume with a center coordinate $\mathbf{w}_\mathrm{NF}^j = [x_\mathrm{NF}^j, y_\mathrm{NF}^j]^T$ projected on the ground, height $H_\mathrm{NF}^j$, $H<H_\mathrm{NF}^j$, $\forall j$, and radius $Q_\mathrm{NF}^{j}$, $\forall j$, cf. Fig. \ref{fig:UAV communication frame}.
With the existence of NFZs, the trajectory of UAV $m$ should satisfy the following inequality in each time slot:\vspace{-2mm}
\begin{equation}\label{eqn:No-Fly area region}
  || \mathbf{q}_m[n] - \mathbf{w}_\mathrm{NF}^j ||^2  \geq \left( Q_\mathrm{NF}^{j} \right)^2,\; \forall n,m,j.
\end{equation}

{\color{black}
\begin{remark}
A reasonable UAV trajectory should be planed such that a UAV can follow the desired path as tightly as possible.
Based on the geometric and kinematics properties, cylindrical volumn models for NFZ satisfy lateral guidance control law of UAVs which are commonly adopted for the design of flight path, e.g. {\cite{Guidance_law_circular_NFZ,Small_UAV_NFZ}}.
When a UAV has to avoid flying over NFZs, a centripetal acceleration on the UAV is generated to help the UAV change its aviation direction.
The centripetal acceleration is related to the UAV's radial flying velocity and radius of turning circle, which leads to a circular trajectory when the UAV turns steadily {\cite{NFZ_Liang}}.
Therefore, for the purpose to design a more practical NFZ model so that a UAV can tightly follow its desired flight path, NFZs are modeled by cylindrical volume constraints in this paper.
\end{remark}
}

\vspace*{-7mm}
\subsection{Channel Model}\vspace*{-2mm}
We assume that the wireless channels from a UAV to the ground user/Eve on each subcarrier are LoS-dominated and we adopt the commonly used free-space path loss model{\color{black}\footnote{{\color{black}We note that {field measurements \cite{Colpaert_2018}} suggest that air-to-ground links are almost guaranteed to be LoS channels in rural areas when a UAV flies with an altitude of 100 meters or above to serve a cell with a radius of 500 meters.
Furthermore, the aviation altitude of a UAV can be adjusted according to the type of terrain and the scale of the cells, which can guarantee that the air-to-ground channel LoS probability approaches one {\cite{Hourani_LAP}}.}}} as in \cite{Zeng_2016_Unmanned_Aerial_Vehicles_Magazine}-\cite{Zeng_2016_Throughput_Maximization_for_UAV}.
Furthermore, we assume that the Doppler effect caused by the mobility of UAVs can be well compensated by all the receivers{\color{black}\footnote{{\color{black}This assumption is commonly used in the research of UAV communications {\cite{Zeng_2017_Energy_Efficient_UAV,Zhong_Cooperative_Jamming,3d_Solar_2019,Propulsion_Energy_Eom,Joint_Altitude_Yang}}, since the impact of frequency offset caused by the high mobility of UAV has been well studied in the literature which can be well compensated by using some Doppler effect compensation technologies {\cite{doppler_Lim,doppler_Jionghui,doppler_Wang}}.}}}.
Thus, the channel power gain from UAV $m$ to ground user $k_U$ in time slot $n$ can be given by\vspace*{-2mm}
\begin{equation}\label{eqn:distance of communication}
  h_{m,{k_U}}[n] = \beta_0 d_{m,{k_U}}^{-2}[n]=\frac{\beta_0}{|| \mathbf{q}_m[n] - \mathbf{w}_{k_U} ||^2 + H^2},\; \forall n, m, k_U,
\end{equation}
\noindent where $\beta_0$ denotes the channel power gain at the reference distance $d_0 = 1$ m.
Similarly, channel power gains from UAV $m$ to Eve $k_E$ in time slot $n$ can be written as\vspace*{-2mm}
\begin{equation}\label{eqn:distance of communication}
  h_{m,{k_E}}[n] = \beta_0 d_{m,{k_E}}^{-2}[n]=\frac{\beta_0}{|| \mathbf{q}_m[n] - \mathbf{w}_{k_E} ||^2 + H^2},\; \forall n, m, k_E.
\end{equation}

\vspace*{-4mm}
\subsection{Communication \& Jamming Scheduling Model}\vspace*{-2mm}
To prevent multiple access interference from all the communication UAVs, in time slot $n$, we assume that there is at most one UAV communicates with at most one user on subcarrier $i$.
On the other hand, since there are multiple Eves randomly distributed on the ground and multiple UAVs in the sky, when subcarrier $i$ is not assigned to UAV $m$ for information transmission, UAV $m$ can act as a jammer on this subcarrier to combat the channel of Eves.
{\color{black} For the sake of presentation and to facilitate the solution design, we denote $s_{m,k_U,i}[n]\in \{0,1\}$ and $s^J_{m,i}[n]\in \{0,1\}$  as binary variables, which indicates that if UAV $m$ communicates with user $k_U$, i.e., $s_{m,k_U,i}[n]=1$ or acts as a jammer, i.e., $s^J_{m,i}[n]=1$ in time slot $n$ on subcarrier $i$, respectively.}
In particular, the binary variables satisfy the following constraints:\vspace*{-3mm}
\begin{align}\label{eqn:int probility_ease}
 s_{m,k_U,i}[n] &\in \{0,1\}, \forall n, m, k_U, i,
  \\
 s^J_{m,i}[n] &\in \{0,1\}, \forall n, m, i,
  \\
 \sum_{m =1}^{M} \sum_{k_U =1}^{K_\mathrm{U}} s_{m,k_U,i}[n] &\le 1, \quad\;\;\; \forall n, i,
  \\
 \sum_{k_U =1}^{K_\mathrm{U}} s_{m,k_U,i}[n] +  s^J_{m,i}[n] &\leq1, \quad\;\;\; \forall n, m, i.
\end{align}

\section{Problem Formulation}
%
In this section, we first define the achievable rate and secrecy rate for the considered multi-UAV system. Then, the multi-UAV secure communication system design is formulated as a non-convex optimization problem.

\vspace*{-4mm}
\subsection{Achievable Rate \& Secrecy Rate}
\vspace*{-1.5mm}
Denote $p_{m,k_U,i}[n] \ge 0$ as the communication power from UAV $m$ to user $k_U$ in time slot $n$ on subcarrier $i$, $p^J_{m,i}[n] \ge 0$ as the jamming power from UAV $m$ in time slot $n$ on subcarrier $i$, and $P^m_\mathrm{peak}$ the peak transmission power of UAV $m$.
Then, we have\vspace*{-2mm}
\begin{align}\label{eqn:max power constrain}
  0 \le \underbrace{\sum_{k_U =1}^{K_\mathrm{U}} \sum_{i=1}^{N_\mathrm{F}} s_{m,k_U,i}[n] p_{m,k_U,i}[n]}_{\mbox{Information power}} + \underbrace{\sum_{i=1}^{N_\mathrm{F}} s^J_{m,i}[n] p^J_{m,i}[n]}_{\mbox{Jamming power}} \le P^m_\mathrm{peak},\; \forall n, m.
\end{align}

Thus, if UAV $m$ is selected to communicate to user $k_U$ in time slot $n$ on subcarrier $i$, i.e., $s_{m,k_U,i}[n] = 1$, the received signal-to-interference-plus-noise ratio (SINR) at user $k_U$ on subcarrier $i$ in time slot $n$ can be written as\vspace*{-2mm}
\begin{equation}\label{eqn:SINR to user}
  \gamma_{m,k_U,i}[n] = \frac{p_{m,k_U,i}[n]h_{m,k_U}[n]}{ I_{m,k_U,i}[n] +\sigma^2},\; \forall  n, m, k_U, i,
\end{equation}
\noindent where $\sigma^2$ denotes the additive white Gaussian noise (AWGN) power at ground users and\vspace*{-2mm}
\begin{align}\label{eqn:communication interference}
  I_{m,k_U,i}[n] =\quad &\sum_{m'=1,m'\neq m}^{M}
  s^J_{m',i}[n] p^J_{m',i}[n] h_{m',k_U}[n], \forall n, m, k_U, i,
\end{align}
represents the co-channel interference caused by UAV $m'\in \mathcal{M}, m'\neq m$, to user $k_U$ on subcarrier $i$.

On the other hand, the received SINR at Eve $k_E$ from UAV $m$ for attempting to decode the signal of user $k_U$ in time slot $n$ on subcarrier $i$ can be written as\vspace*{-2mm}
\begin{equation}\label{eqn:SINR to Eve}
  \gamma'_{m,k_U,k_E,i}[n] = \frac{p_{m,k_U,i}[n]h_{m,k_E}[n]}{ I_{m,k_E,i}[n] +\sigma^2},\; \forall n, m, k_U, k_E, i,
\end{equation}
\vspace*{-3mm}where\vspace*{-2.5mm}
\begin{align}\label{eqn:communication interference}
  I_{m,k_E,i}[n] = \sum_{m'=1,m'\neq m}^{M}  s^J_{m',i}[n] p^J_{m',i}[n]h_{m',k_E}[n], \forall n, m, k_E, i,
\end{align}
is the interference to Eve $k_E$ on subcarrier $i$ in time slot $n$.

Thus, the communication rate $R_{m,k_U,i}[n]$ from UAV $m$ to user $k_U$ in time slot $n$ on subcarrier $i$ in bits/second/Hertz (bps/Hz) is given by\vspace*{-2mm}
\begin{equation}\label{eqn:communication rate}
  R_{m,k_U,i}[n] = s_{m,k_U,i}[n] \log_2\left(1 + \gamma_{m,k_U,i}[n] \right),\; \forall n, m, k_U, i,
\end{equation}
while the corresponding leakage rate $R'_{m,k_E,i}[n]$ to Eve $k_E$ can be written as\vspace*{-2mm}
\begin{equation}\label{eqn:leakage rate}
  R_{m,k_U,k_E,i}'[n] =   s_{m,k_U,i}[n] \log_2\left(1 + \gamma'_{m,k_U,k_E,i}[n] \right),\; \forall n, m, k_U, k_E, i.
\end{equation}

Then, the average secrecy rate $\bar R^s_{k_U}$ in bps/Hz over $N$ time slots for user $k_U$ considered with $K_E$ eavesdroppers is given by
\begin{align}\label{eqn:distance of communication}
   \bar R^s_{k_U} = \frac{1}{N} \sum_{n=1}^{N}  \sum_{m=1}^{M}  \sum_{i=1}^{N_\mathrm{F}} \bigg [ R_{m,k_U,i}[n] -   \underset{k_E \in \mathcal{K}_{E}}\max\left\{R'_{m,k_U,k_E,i}[n]\right\}  \bigg ] ^+, \forall k_U.
\end{align}

\subsection{Optimization Problem Formulation}
\vspace*{-1.5mm}
For notational simplicity, we denote $\mathcal{Q} = \{\mathbf{q}_m[n], \forall n, m\}$ as the set of all UAVs' trajectory variables,
$ \mathcal{S}_U = \{s_{m,k_U,i}[n] , \forall n,m,k_U,i\}$ as the set of the communication scheduling variables,
$ \mathcal{S}_J =  \{s^J_{m,i}[n], \forall n,m,i\}$ as the set of the jamming scheduling variables.
Let $ \mathcal{P}_U  =  \{p_{m,k_U,i}[n], \forall n, m, k_U, i\}$
denote the set of all the communication UAVs' transmit power variables
and $ \mathcal{P}_J = \{p^J_{m,i}[n] , \forall n, m, i\}$ denote the set of the jamming power variables from all jamming UAVs.

We aim to maximize the average minimum secrecy rate among all ground users via jointly optimizing communication scheduling, communication power, jamming scheduling, jamming power, and all UAVs' trajectories.
Define $\eta$ as an auxiliary optimization variable and the max-min fairness optimization problem can be formulated as\vspace*{-2mm}
\begin{align}\label{eqn:problem 1}
\underset{\eta, \mathcal{Q}, \mathcal{S}_U, \mathcal{S}_J, \mathcal{P}_U, \mathcal{P}_J}{\maxo}\,\,
&  \eta \\
  \mathrm{s.t.}\,
  \mathrm{C1}:\; &\frac{1}{N} \sum_{n=1}^{N}  \sum_{m=1}^{M} \sum_{i=1}^{N_\mathrm{F}} \bigg [ R_{m,k_U,i}[n] -  R'_{m,k_U,k_E,i}[n] \bigg ] ^+ \geq \eta, \forall k_U,k_E, \notag
  \\
  \mathrm{C2}:\;  &s_{m,k_U,i}[n]\in \{0,1\}, \forall n, m, k_U, i, \notag
  \\
  \mathrm{C3}:\;  &s^J_{m,i}[n]\in \{0,1\}, \forall n, m, i,\notag
  \\
  \mathrm{C4}:\;  &\sum_{m =1}^{M} \sum_{k_U =1}^{K_\mathrm{U}}  s_{m,k_U,i}[n] \le 1, \forall n, i,\notag
  \\
  \mathrm{C5}:\;  &\sum_{k_U =1}^{K_\mathrm{U}} s_{m,k_U,i}[n] +  s^J_{m,i}[n] \leq1, \forall n, m, i, \notag
  \\
  \mathrm{C6a}:\;  & \sum_{k_U =1}^{K_\mathrm{U}} \sum_{i=1}^{N_\mathrm{F}} s_{m,k_U,i}[n] p_{m,k_U,i}[n] + \sum_{i=1}^{N_\mathrm{F}} s^J_{m,i}[n] p^J_{m,i}[n]\le P^m_\mathrm{peak},\; \forall n, m,\notag
  \\
  \mathrm{C6b}:\;  & p_{m,k_U,i}[n] \ge 0, \forall n,m,k_U,i, \quad\quad
  \mathrm{C6c}:\;    p^J_{m,i}[n]   \ge 0, \forall n,m,i,\notag
  \\
  \mathrm{C7}:\;  &|| \mathbf{q}_m[n] - \mathbf{q}_m\left[n-1\right] ||^2 \leq V^2, \forall n,m, \notag\\
  \mathrm{C8}:\;  &|| \mathbf{q}_m[n] - \mathbf{w}_\mathrm{NF}^j ||^2  \geq \left( Q_\mathrm{NF}^{j} \right)^2, \forall n, m, j, \notag
  \\
  \mathrm{C9}:\;  &|| \mathbf{q}_m[n] - \mathbf{q}_{m'}\left[n\right] ||^2 \geq \hspace*{-1mm} D_\mathrm{S}^2, \forall n,m, m \hspace*{-1mm} \neq \hspace*{-1mm} m',  \notag
  \\
  \mathrm{C10}:\;  &\mathbf{q}_{m}[0] = \mathbf{q}_{m}^0, \forall m, \quad\quad\quad\quad\quad\quad\;
  \mathrm{C11}:\;  \mathbf{q}_{m}[N]  = \mathbf{q}_{m}^F, \forall m,\notag
\end{align}
\noindent where constraint $\mathrm{C1}$ is imposed to guarantee an average minimum secrecy rate $\eta$ for each user over all time slots.
$\mathrm{C2}$ and $\mathrm{C3}$ are the binary constraints to denote the UAV's communication and jamming subcarrier allocation, respectively.
Constraint $\mathrm{C4}$ is imposed to make sure that in any time slot, each subcarrier can be allocated to at most one user for communication from all the UAVs.
Constraint $\mathrm{C5}$ is introduced such that a UAV can either transmit information or jamming signal on each subcarrier.
Constraints $\mathrm{C6a}$$-$$\mathrm{C6c}$ are the UAV transmission power constraints, $P^m_\mathrm{peak}$ in $\mathrm{C6a}$ denotes the maximum transmission power of UAV $m$, $\mathrm{C6b}$ and $\mathrm{C6c}$ are the non-negative constraints on the power allocation variables.
Constraint $\mathrm{C7}$ is imposed to ensure that each UAV should fly no faster than its maximum speed $V$ in each time slot.
Constraint $\mathrm{C8}$ states that a UAV is prohibited to fly over the NFZs.
$D_\mathrm{S}$ in constraint $\mathrm{C9}$ is the minimum distance between any UAV pairs to avoid collision.
Constraints $\mathrm{C10}$ and $\mathrm{C11}$ indicate the fixed initial and final locations of the UAVs, $\mathbf{q}_{m}^0 = [x_m[0],y_m[0]]^T$ and $\mathbf{q}_{m}^F = [x_m[N],y_m[N]]^T$, respectively.
%

\vspace*{3mm}
\section{Joint Trajectory and Resource Allocation Design}
\vspace*{1.5mm}
%

\begin{figure}[t]
  \centering
  \includegraphics[width=4.9 in]{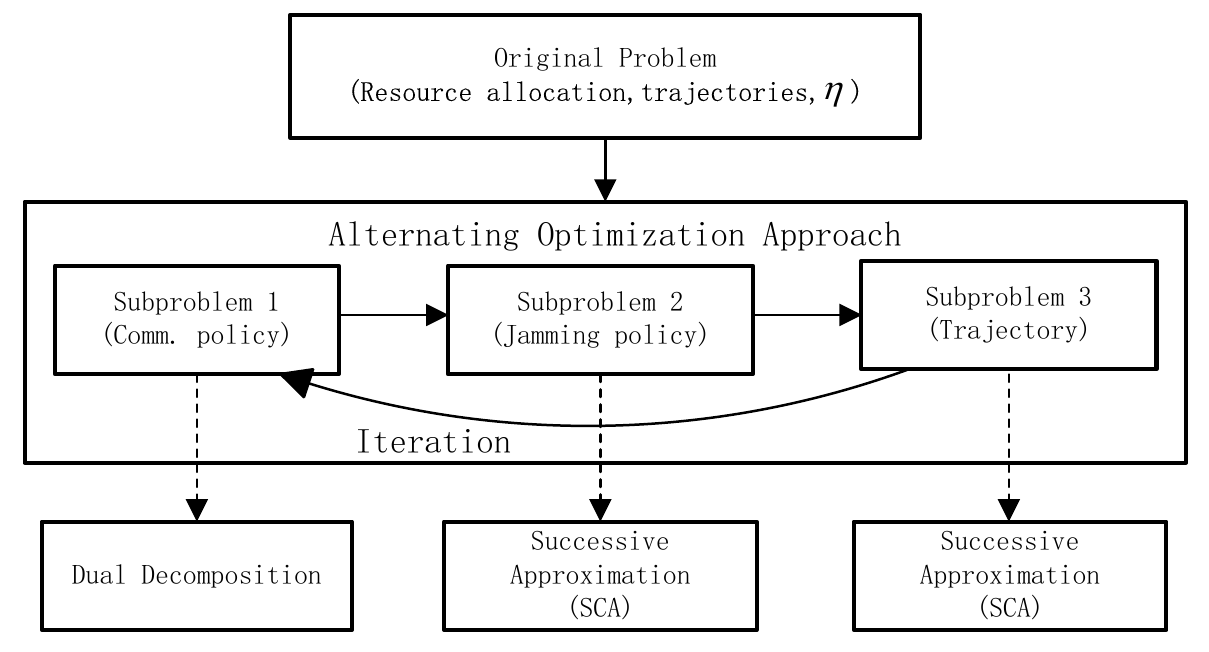}
  \vspace*{-2mm}
  \caption{A flow chart of the proposed iterative algorithm.}
  \vspace*{-3mm}
  \label{fig:iterative alternating algorithm}
\end{figure}

The formulated problem in \eqref{eqn:problem 1} is a mixed-integer non-convex and generally intractable.
Specifically, the main obstacle in solving \eqref{eqn:problem 1} arises from the binary constraints $\mathrm{C2}$ and $\mathrm{C3}$ while the non-convexity originates from $\mathrm{C1}$, $\mathrm{C6a}$, $\mathrm{C8}$, and $\mathrm{C9}$.
Besides, the problem is further complicated by the coupling between binary and continuous variables in $\mathrm{C1}$ and $\mathrm{C6a}$.
Therefore, obtaining the globally optimal solution requires a prohibitively large computational complexity which is not practical even for moderate system size.
As a compromise approach, in this section, we propose a series of transformation which facilitates the development of a low-complexity algorithm for achieving a suboptimal solution{\color{black}\footnote{{\color{black}In fact, the designed suboptimal algorithm requires only a polynomial time computational complexity which is suitable for realtime implementations {\cite{Cormen_2001}}.
In the considered system, the algorithm can be executed by the UAV-onboard computing system or with the help of a ground station via computation offloading {\cite{nguyen2018introduction}}.}}}.
To this end, we first propose Lemma \ref{lemma1} to handle the non-smoothness function due to the $[\cdot ]^+$ operation in constraint $\mathrm{C1}$ of problem \eqref{eqn:problem 1}.\vspace*{2mm}

\begin{Lem}\label{lemma1}
Problem \eqref{eqn:problem 1} has the same optimal solution as the following problem:
%
\begin{align}\label{eqn:no nonsm problem 1}
\underset{\eta, \mathcal{Q}, \mathcal{S}_U, \mathcal{S}_J, \mathcal{P}_U, \mathcal{P}_J}{\maxo}\,\,\eta \\
  \mathrm{s.t.}\quad  \mathrm{C2} - &\mathrm{C11}, \notag\\
  \mathrm{C1}:\, &\frac{1}{N} \sum_{n=1}^{N}  \sum_{m=1}^{M}  \sum_{i=1}^{N_\mathrm{F}}  R_{m,k_U,i}[n] -  R'_{m,k_U,k_E,i}[n] \geq \eta, \forall k_U,k_E. \notag
\end{align}
\end{Lem}
\vspace*{1.5mm}
\begin{Proof}
Please refer to Appendix A for a proof of Lemma \ref{lemma1}.
\end{Proof}
\vspace*{2.5mm}
Although problem \eqref{eqn:no nonsm problem 1} is easier to handle, it is still non-convex and challenging to solve.
Next, we divide problem \eqref{eqn:no nonsm problem 1} into three subproblems with three individual sets of optimization variables, respectively, i.e., $(\eta, {\mathcal{S}_U},{\mathcal{P}_U})$, $(\eta, {\mathcal{S}_J},{\mathcal{P}_J})$, $(\eta, \mathcal{Q})$, which facilitates the design of a computationally efficient iterative alternating algorithm to achieve a suboptimal solution, cf. Fig. \ref{fig:iterative alternating algorithm}.
In particular, we first jointly optimize the average minimum secrecy data rate, communication user scheduling, and communication power of each UAV.
Then, with fixed trajectories and obtained communication policy, we study the jamming scheduling and jamming power allocation, for obtaining an intermediate optimized average minimum secrecy data rate.
In the last subproblem, we design all UAVs' trajectories and update the average minimum secrecy data rate with the above obtained resource allocation policy.

\subsection{Subproblem 1: Communication Resource Allocation Optimization}
%
In this section, we consider subproblem 1 for optimizing the communication user scheduling and communication power by assuming that jamming policy $({\mathcal{S}_J}, {\mathcal{P}_J})$ and all UAVs' trajectories $\mathcal{Q}$ are fixed.
Thus, subproblem 1 can be written as:\vspace*{-2mm}
\begin{align}\label{eqn:subproblem 1}
\underset{\eta,\, \mathcal{S}_U, \mathcal{P}_U}{\maxo}\,\,
&  \eta \\
  \mathrm{s.t.}\,
  \mathrm{C1}:\; &\frac{1}{N} \sum_{n=1}^{N}  \sum_{m=1}^{M}  \sum_{i=1}^{N_\mathrm{F}}  R_{m,k_U,i}[n] -  R'_{m,k_U,k_E,i}[n]  \geq \eta, \forall k_U,k_E, \notag
  \\
  \mathrm{C2}:\;  &s_{m,k_U,i}[n]\in \{0,1\}, \forall n, m, k_U, i,
  \quad
  \mathrm{C4}:\;  \sum_{m =1}^{M} \sum_{k_U =1}^{K_\mathrm{U}}  s_{m,k_U,i}[n] \le 1, \forall n, i,\notag
  \\
  \mathrm{C5}:\;  &\sum_{k_U =1}^{K_\mathrm{U}} s_{m,k_U,i}[n] +  s^J_{m,i}[n] \leq1, \forall n, m, i, \notag
  \\
  \mathrm{C6a}:\;  &  \sum_{k_U =1}^{K_\mathrm{U}} \sum_{i=1}^{N_\mathrm{F}} s_{m,k_U,i}[n] p_{m,k_U,i}[n] + \sum_{i=1}^{N_\mathrm{F}} s^J_{m,i}[n] p^J_{m,i}[n]\le P^m_\mathrm{peak},\; \forall n, m,\notag
  \\
  \mathrm{C6b}:\;  & p_{m,k_U,i}[n] \ge 0, \forall n,m,k_U,i.\notag
\end{align}
%
%
In order to solve subproblem 1 in \eqref{eqn:subproblem 1}, by following a similar approach as \cite{Yuanxin_Cai}, \cite{Ng_L}, we introduce an auxiliary variable $\widetilde{p}_{m,k_U,i}[n] = s_{m,k_U,i}[n] p_{m,k_U,i}[n]$, and the problem can be equivalently written as\vspace*{-2mm}
\begin{align}\label{eqn:problem 1 tilde p}
\underset{\eta,\, \mathcal{S}_U, \widetilde{\mathcal{P}}_U}{\maxo}\,\,
&  \eta \\
  \mathrm{s.t.}\,
  \widetilde{\mathrm{C1}}:\; &\frac{1}{N} \sum_{n=1}^{N}  \sum_{m=1}^{M}  \sum_{i=1}^{N_\mathrm{F}}  \widetilde{R}_{m,k_U,i}[n] - \widetilde{R}'_{m,k_U,k_E,i}[n] \geq \eta, \forall k_U,k_E, \notag
  \\
  \mathrm{C2}:\;  &s_{m,k_U,i}[n]\in \{0,1\}, \forall n, m, k_U, i,
  \quad
  \mathrm{C4}:\;  \sum_{m =1}^{M} \sum_{k_U =1}^{K_\mathrm{U}} s_{m,k_U,i}[n] \le 1, \forall n, i,\notag
  \\
  \mathrm{C5}:\;  &\sum_{k_U =1}^{K_\mathrm{U}} s_{m,k_U,i}[n] +  s^J_{m,i}[n] \leq1, \forall n, m, i, \notag
  \\
  \widetilde{\mathrm{C6a}}:\;  &  \sum_{k_U =1}^{K_\mathrm{U}} \sum_{i=1}^{N_\mathrm{F}}\widetilde{p}_{m,k_U,i}[n] + \sum_{i=1}^{N_\mathrm{F}}s^J_{m,i}[n] p^J_{m,i}[n]\le P^m_\mathrm{peak},\; \forall n, m,\notag
  \\
  \widetilde{\mathrm{C6b}}:\;  &\widetilde{p}_{m,k_U,i}[n] \ge 0, \forall n,m,k_U,i, \notag
\end{align}
where $\widetilde{\mathcal{P}}_U = \{ \widetilde{p}_{m,k_U,i}[n], \forall  n, m, i,  k_U\}$,\vspace*{-2mm}
\begin{align}\label{eqn:problem 1 tilde p more detail}
\widetilde{R}_{m,k_U,i}[n]  &= s_{m,k_U,i}[n] \log_2\left(1 +  \frac{\widetilde{p}_{m,k_U,i}[n] \mathcal{H}_{m,k_U,i}[n] }{s_{m,k_U,i}[n] }  \right), \forall n, m, k_U, i, \\
\widetilde{R}'_{m,k_U,k_E,i}[n] &= s_{m,k_U,i}[n] \log_2\left(1  +  \frac{\widetilde{p}_{m,k_U,i}[n] \mathcal{H}'_{m,k_E,i}[n] }{s_{m,k_U,i}[n] }  \right), \forall n, m, k_U, k_E, i, \\
\mathcal{H}_{m,k_U,i}[n] &=   \frac{h_{m,k_U}[n]}{ I_{m,k_U,i}[n] +\sigma^2}, \forall n, m, k_U, i, \\
\mathcal{H}'_{m,k_E,i}[n] &= \frac{h_{m,k_E}[n]}{ I_{m,k_E,i}[n] +\sigma^2}, \forall n, m, k_E, i.
\end{align}

Next, we handle the binary user scheduling constraint $\mathrm{C2}$ in \eqref{eqn:problem 1 tilde p}. In particular, we follow a similar approach as in \cite{li,Ng_L} and relax the binary subcarrier variable $s_{m,k_U,i}[n]$ as a real value between 0 and 1, i.e.,\vspace*{-2mm}
\begin{equation}
\widetilde{\mathrm{C2}}: 0 \le s_{m,k_U,i}[n] \le 1, \forall n, m, k_U, i.
\end{equation}
Meanwhile, the relaxed variable $s_{m,k_U,i}[n]$ serves as a time-sharing factor for user $k_U$ on subcarrier $i$ in time slot $n$.

After replacing $\mathrm{C2}$ with $\widetilde{\mathrm{C2}}$ in \eqref{eqn:problem 1 tilde p}, the problem can be written\footnote{We note that the adopted constraint relaxation is tight as will be shown in the following.} as\vspace*{-2mm}
\begin{align}\label{eqn:problem 1 relax s}
\underset{\eta,\, \mathcal{S}_U,\tilde{\mathcal{P}}_U}{\maxo}\,\,
& \eta  \\
  \mathrm{s.t.}\quad \;
  &\widetilde{\mathrm{C1}}, \mathrm{C4} - \widetilde{\mathrm{C6b}}, \notag
  \\
  &\widetilde{\mathrm{C2}}:\;  0 \le s_{m,k_U,i}[n] \le 1, \forall n, m, k_U, i. \notag
\end{align}

Then, before we derive the optimal communication resource allocation, we first verify the convexity of constraint $\widetilde{\mathrm{C1}}$ jointly with respect to (w.r.t.) $\widetilde{s}_{m,k_U,i}[n]$ and $\widetilde{p}_{m,k_U,i}[n]$ via the following lemma:

\begin{Lem}\label{lemma2}
{\color{black}
For $\kappa_1 > \kappa_2\ge0$, the function $\psi(x, y) \triangleq x\log_2(1 + \frac{\kappa_1 y}{x}) - x \log_2(1 + \frac{\kappa_2 y}{x})$ is concave w.r.t. $x\ge0$ and $y \ge 0$.
}
\end{Lem}
\begin{Proof}
Please refer to Appendix B for a proof of Lemma \ref{lemma2}.
\end{Proof}

In other words, with $\mathcal{H}_{m,k_U,i}[n] > \mathcal{H}'_{m,k_E,i}[n]$, constraint $\widetilde{\mathrm{C1}}$ is jointly concave w.r.t. $\mathcal{S}_U$ and $\widetilde{\mathcal{P}}_U$ which satisfies Lemma 2.
Next, we consider the communication resource allocation under $\mathcal{H}_{m,k_U,i}[n] > \mathcal{H}'_{m,k_E,i}[n]$, and study the convexity of constraint $\widetilde{\mathrm{C1}}$ under such a condition.
\begin{Lem}\label{lemma3}
For the problem in \eqref{eqn:problem 1 tilde p}, if $\mathcal{H}_{m,k_U,i}[n] \le \mathcal{H}'_{m,k_E,i}[n]$, UAV $m$ would not allocate any power for user $k_U$ on subcarrier $i$ in time slot $n$, i.e., $s_{m,k_U,i}[n] = 0$ and $\widetilde{p}_{m,k_U,i}[n] = 0$.
\end{Lem}\vspace*{-1mm}
\begin{Proof}
Please refer to Appendix C for a proof of Lemma 3.
\end{Proof}

Now, problem \eqref{eqn:problem 1 relax s} is jointly convex w.r.t. $\eta, \widetilde{s}_{m,k_U,i}[n]$, and $\widetilde{p}_{m,k_U,i}[n]$.
Furthermore, problem \eqref{eqn:problem 1 relax s} satisfies the Slater's constraint qualification and thus the strong duality holds \cite{Boyd,sun_yan_tcom_noma,channel_power_gain_Gopala}.
Therefore, the duality gap is zero.
In other words, the optimal solution of problem \eqref{eqn:problem 1 relax s} can be obtained by solving its dual problem.
To shed lights on important system design insights, we solve the dual problem via deriving some semi-closed-form solutions.
To this end, we first derive the Lagrangian of problem \eqref{eqn:problem 1 relax s}:\vspace*{-2mm}
\begin{align}\label{eqn:subproblem_1_Lag}
&{\cal L}(\eta, {\bm\alpha} ,{\bm\beta}, {\bm\varepsilon}, {\bm\vartheta}, \mathcal{S}_U, \tilde{\mathcal{P}}_U) \\
&=\hspace{-0.05in}
\eta \hspace{-0.05in} - \hspace{-0.12in} \sum\limits_{{k_U} = 1}^{{K_\mathrm{U}}} \hspace{-0.05in} \sum_{k_E=1}^{K_\mathrm{E}} \hspace{-0.05in} \alpha _{{k_U,k_E}} \hspace{-0.05in} \left( \hspace{-0.08in} N\eta
\hspace{-0.05in} - \hspace{-0.12in}
\sum\limits_{n = 1}^N \hspace{-0.04in} \sum\limits_{m = 1}^M  \hspace{-0.04in} \sum_{i=1}^{N_\mathrm{F}} \hspace{-0.04in} \left[ {{{\tilde R}_{m,{k_U},i}}[n]
\hspace{-0.05in} - \hspace{-0.05in}
\widetilde{R}'_{m,k_U,k_E,i}[n] } \right]     \hspace{-0.05in} \right)
\hspace{-0.08in} - \hspace{-0.1in}
\sum\limits_{n = 1}^N  \hspace{-0.04in} \sum_{i=1}^{N_\mathrm{F}}\hspace{-0.05in} \beta_{i}[n]
\hspace{-0.05in} \left( \hspace{-0.05in} \sum_{m=1}^M \hspace{-0.04in} \sum_{k_U=1}^{{K}_\mathrm{U}} \hspace{-0.05in} s_{m,k_U,i}[n] \hspace{-0.05in} - \hspace{-0.05in} 1 \hspace{-0.08in} \right)
\notag\\
&\;\;- \hspace{-0.05in} \sum\limits_{n\hspace{-0.01in} =\hspace{-0.01in} 1}^N \hspace{-0.04in} \sum\limits_{m  \hspace{-0.01in}= \hspace{-0.01in} 1}^M \hspace{-0.04in}
\left[\hspace{-0.04in}
\sum\limits_{i \hspace{-0.01in} = \hspace{-0.01in} 1}^{{N_{\rm{F}}}} \hspace{-0.04in} {\varepsilon _{m,i}}\hspace{-0.04in} \left[ n \right]
\hspace{-0.06in} \left( \hspace{-0.04in} {\sum\limits_{{k\hspace{-0.01in}_U} \hspace{-0.01in} = \hspace{-0.01in} 1}^{{K_\mathrm{U}}} \hspace{-0.08in} {{s_{m,{k_U},i}}[n]} \hspace{-0.04in} + \hspace{-0.04in} s_{m,i}^J \hspace{-0.04in} \left[ n \right] \hspace{-0.06in} - \hspace{-0.06in} 1} \hspace{-0.08in} \right)
\hspace{-0.06in}-\hspace{-0.06in}
{\vartheta _m} \hspace{-0.04in} \left[ n \right]
\hspace{-0.06in}  \left( \hspace{-0.04in} \sum\limits_{{k\hspace{-0.01in}_U}\hspace{-0.01in} =\hspace{-0.01in} 1}^{{K_\mathrm{U}}}\hspace{-0.04in}  \sum_{i\hspace{-0.01in}=\hspace{-0.01in}1}^{N_\mathrm{F}} \hspace{-0.04in} {{\tilde p}_{m\hspace{-0.01in},{k_U}\hspace{-0.01in},i}} \hspace{-0.02in} [n]
\hspace{-0.06in} + \hspace{-0.06in} \sum_{i=1}^{N_\mathrm{F}} \hspace{-0.04in} {s_{m,i}^Jp_{m,i}^J \hspace{-0.02in} [n]} \hspace{-0.06in}  - \hspace{-0.06in} P\hspace{-0.01in}_\mathrm{peak}^m \hspace{-0.06in} \right) \hspace{-0.04in} \right] , \notag
\end{align}
where ${\bm\alpha} = \{ \alpha_{k_U,k_E} \ge 0, \forall k_U,k_E \}$, ${\bm\beta} = \{ \beta_{i}[n] \ge 0, \forall n,i \}$, ${\bm\varepsilon} = \{ \varepsilon_{m,i}[n] \ge 0, \forall n,m,i \}$, and ${\bm\vartheta} = \{ \vartheta_m[n] > 0, \forall n,m \}$, denote the Lagrange multipliers for constraints $\widetilde{\mathrm{C1}}$, ${\mathrm{C4}}$, ${\mathrm{C5}}$, and $\widetilde{\mathrm{C6a}}$, respectively.
Constraints $\mathrm{C2}$ and $\widetilde{\mathrm{C6b}}$ will be considered when deriving the optimal solution via examining the Karush-Kuhn-Tucker (KKT) conditions in the following.
Thus, the dual problem of \eqref{eqn:problem 1 relax s} can be written as\vspace*{-2mm}
\begin{align}\label{eqn:subproblem_1_dual problem}
\mathcal{D} = \underset{{\bm\alpha} ,{\bm\beta}, {\bm\varepsilon}, {\bm\vartheta} \ge 0}{\mino} \; \underset{\eta,\, \mathcal{S}_U, \tilde{\mathcal{P}}_U}{\maxo}\; {\cal L}(\eta, {\bm\alpha} ,{\bm\beta}, {\bm\varepsilon}, {\bm\vartheta}, \mathcal{S}_U, \tilde{\mathcal{P}}_U).
\end{align}

Then, by using dual decomposition \cite{Derrick_Secure_OFDMA}, the dual problem can be solved iteratively by solving the two layers which is divided from the dual problem:
Layer 1, maximizing the Lagrangian over minimum secrecy rate $\eta$, user scheduling $\mathcal{S}_U$, and power allocation $\tilde{\mathcal{P}}_U$ in \eqref{eqn:subproblem_1_dual problem}, for given Lagrange multipliers ${\bm\alpha}, {\bm\beta}, {\bm\varepsilon}$, and ${\bm\vartheta}$;
Layer 2, minimizing the Lagrangian function over ${\bm\alpha}, {\bm\beta}, {\bm\varepsilon}$, and ${\bm\vartheta}$ in \eqref{eqn:subproblem_1_dual problem}, for a fixed minimum secrecy rate $\eta$, user scheduling $\mathcal{S}_U$, and power allocation $\tilde{\mathcal{P}}_U$.

\emph{Solution of Layer 1 (Power Allocation and User Scheduling):}
Denote $s^*_{m,k_U,i}[n]$, ${p}^*_{m,k_U,i}[n]$, and $\widetilde{p}^*_{m,k_U,i}[n]$ the optimal solutions of subproblem 1.
Thus, the optimal power allocation for user $k_U$ on subcarrier $i$ in time slot $n$ is given by\vspace*{-2mm}
\begin{align}\label{eqn:subproblem_1_optimal power}
&\widetilde{p}^*_{m,k_U,i}[n] = s_{m,k_U,i}[n] {p}^*_{m,k_U,i}[n]
\\
= & \frac{s_{m\hspace{-0.01in},k_U\hspace{-0.01in},i}\hspace{-0.01in}[n]}{2}
\left[ \sqrt{ \Gamma_{m,k_U,k_E,i}^2[n] + \frac{4\alpha_{k_U,k_E}}{\vartheta_m[n] \ln2} \Gamma_{m,k_U,k_E,i}[n] }
-
\left( \frac{1}{\mathcal{H}'_{m,k_E,i}[n]} + \frac{1}{\mathcal{H}_{m,k_U,i}[n]} \right)
\right]^{\hspace{-0.04in}+}  \hspace{-0.03in},\notag
\end{align}
where $\Gamma_{m,k_U,k_E,i}[n] = \left( \frac{1}{\mathcal{H}'_{m,k_E,i}[n]} - \frac{1}{\mathcal{H}_{m,k_U,i}[n]} \right)$.

We note that the solution derived in \eqref{eqn:subproblem_1_optimal power} coincides Lemma 3 where no power is allocated on subcarrier $i$ from UAV $m$ if $\mathcal{H}_{m,k_U,i}[n] \le \mathcal{H}'_{m,k_E,i}[n]$.
Lagrange multipliers $\alpha_{k_U,k_E}$ and $\vartheta_m[n]$ in \eqref{eqn:subproblem_1_optimal power} ensure that the average minimum secrecy rate constraint $\widetilde{\mathrm{C1}}$ and the maximum transmission power constraint $\widetilde{\mathrm{C6a}}$ are satisfied, respectively, when the optimal solution of \eqref{eqn:problem 1 relax s} is attained.
In general, the water-level for user $k_U$, i.e. $\frac{4\alpha_{k_U,k_E}}{\vartheta_m[n] \ln2}\Gamma_{m,k_U,k_E,i}[n]$, is different from other users on subcarrier $i$ in time slot $n$.
According to the KKT conditions \cite{Boyd}, the following equality holds at the optimal point of the problem in \eqref{eqn:problem 1 tilde p}:\vspace*{-2mm}
\begin{equation}\label{eqn:problem 1.1_Dual}
\alpha_{k_U,k_E}\left(  \sum_{n=1}^{N}  \sum_{m=1}^{M}  \sum_{i=1}^{N_\mathrm{F}} \bigg [ R_{m,k_U,i}[n]  -   \widetilde{R}'_{m,k_U,k_E,i}[n]   \bigg ] -  N\eta
 \right) = 0, \forall k_U,k_E.\vspace*{-1mm}
\end{equation}
Therefore, Lagrange multiplier $\alpha_{k_U,k_E}$ is used to adjust the resource allocation such that constraint $\mathrm{C1}$ is satisfied with equality.
In fact, it reallocates the resource from the stronger users to weaker users to achieve certain fairness between users.
On the other hand, Lagrange multiplier $\vartheta_m[n] > 0$ adjusts the water level to satisfy constraint $\widetilde{\mathrm{C6a}}$.
Then, the optimal subcarrier allocation can be obtained via the derivative of the Lagrangian function w.r.t. $s_{m,k_U,i}[n]$, which yields
\begin{align}\label{eqn:subproblem_1_optimal scheduling}
S\hspace{-0.02in}_{m\hspace{-0.01in},k_U\hspace{-0.02in},i}\hspace{-0.02in}[n]
\hspace{-0.05in} = \hspace{-0.05in}
\alpha\hspace{-0.01in}_{k\hspace{-0.01in}_U,k\hspace{-0.01in}_E} \hspace{-0.05in}
\left( \hspace{-0.05in} {\log _2} \hspace{-0.05in}
\left( \hspace{-0.04in} {\frac{{1 \hspace{-0.05in} + \hspace{-0.05in} {\Lambda _{m\hspace{-0.01in},{k\hspace{-0.01in}_U}\hspace{-0.01in},i}}[n]}}{{1 \hspace{-0.05in} + \hspace{-0.05in} \Lambda {'_{m\hspace{-0.01in},{k\hspace{-0.01in}_U}\hspace{-0.01in},
k\hspace{-0.01in}_E\hspace{-0.01in},i}}\hspace{-0.01in}[n]}}}\hspace{-0.04in} \right)
\hspace{-0.05in} - \hspace{-0.05in}
\frac{{ {\Lambda _{m,{k_U},i}}[n]}}{{\left( \hspace{-0.02in} {1 \hspace{-0.05in} + \hspace{-0.05in} {\Lambda _{m\hspace{-0.01in},{k\hspace{-0.01in}_U}\hspace{-0.01in},i}}\hspace{-0.01in}[n]} \hspace{-0.02in} \right) \hspace{-0.04in} \ln 2}}
\hspace{-0.05in} + \hspace{-0.05in}
\frac{{ \Lambda {'_{m,{k_U},k_E,i}}[n]}}{{\left( \hspace{-0.02in} {1 \hspace{-0.05in} + \hspace{-0.05in} \Lambda {'_{m\hspace{-0.01in},{k\hspace{-0.01in}_U}\hspace{-0.01in},
k\hspace{-0.01in}_E\hspace{-0.01in},i}}\hspace{-0.01in}[n]} \right)\hspace{-0.04in}\ln 2}} \hspace{-0.04in} \right)
\hspace{-0.06in} - \hspace{-0.05in}
\beta\hspace{-0.01in}_i\hspace{-0.01in}[n] \hspace{-0.05in} - \hspace{-0.05in} \varepsilon\hspace{-0.01in}_{m\hspace{-0.01in},i}\hspace{-0.01in}[n],
\end{align}
where ${\Lambda _{m,{k_U},i}}[n] = {p}_{m,k_U,i}[n]  \mathcal{H}_{m,k_U,i}[n]$ and ${\Lambda '_{m,{k_U},k_E,i}}[n] = p_{m,k_U,i}[n] {\mathcal{H}}'_{m,k_E,i}[n]$.
Since \eqref{eqn:subproblem_1_optimal scheduling} is independent of $s_{m,k_U,i}[n]$, with the consideration of constraint ${\mathrm{C4}}$, the optimal user scheduling on subcarrier $i$ for UAV $m$ in each time slot $n$ is given by\vspace*{-2mm}
\begin{eqnarray} \label{eqn:subproblem_1_optimal scheduling policy}
s^*_{m,k_U,i}[n]= \left\{
  \begin{array}{ll}
  1, & m^*,k_U^*=\underset{m,k_U} \max (S_{m,k_U,i}[n]), \\
  0, & \mathrm{otherwise},
  \end{array}
\right.
\forall n,i,
\end{eqnarray}
the solution is still binary, which means that the relaxation adopted in $\widetilde{\mathrm{C2}}$ is tight.

\emph{Solution of Layer 2 (Master Problem):}
To solve the master minimization problem in \eqref{eqn:subproblem_1_dual problem}, we adopt the gradient method to update the Lagrange multipliers which is given by\vspace*{-1mm}
\begin{eqnarray}
\hspace{-0.2in} \alpha_{k_U,k_E} \hspace{-0.02in} (l_1 \hspace{-0.05in} + \hspace{-0.05in} 1)\hspace{-0.15in} &=& \hspace{-0.15in} \bigg[  \hspace{-0.02in} \alpha_{k_U,k_E}(l_1) \hspace{-0.05in}  - \hspace{-0.05in}  \delta_1(l_1) \hspace{-0.05in} \times \hspace{-0.05in}
 \bigg(
  \frac{1}{N}\hspace{-0.05in} \sum_{n=1}^{N}  \sum_{m=1}^{M} \sum_{i=1}^{N_\mathrm{F}} \hspace{-0.04in} \bigg [ \hspace{-0.04in} R_{m,k_U,i}[n] \hspace{-0.05in} - \hspace{-0.05in} R'_{m,k_U,k_E,i}[n] \hspace{-0.02in} \bigg ] \hspace{-0.08in} - \hspace{-0.05in} \eta
  \bigg)\hspace{-0.05in} \bigg]^{\hspace{-0.02in}+}\hspace{-0.08in} ,  \forall k_U,k_E, \label{eqn:alpha} \\
\hspace{-0.2in} \beta_{i}[n](l_1 \hspace{-0.05in} + \hspace{-0.05in} 1)\hspace{-0.15in} &=&\hspace{-0.15in} \bigg[ \beta_{i}[n](l_1) \hspace{-0.05in} - \hspace{-0.05in} \delta_2(l_1) \times
\bigg(
 1 - \sum_{m=1}^M \sum\limits_{{k_U} = 1}^{{K_\mathrm{U}}} s_{m,k_U,i}[n]
\bigg) \bigg]^+  , \forall n, i, \label{eqn:beta} \\
\hspace{-0.2in} \varepsilon_{m,i}[n](l_1 \hspace{-0.05in} + \hspace{-0.05in} 1)\hspace{-0.15in} &=&\hspace{-0.15in} \bigg[ \varepsilon_{m,i}[n](l_1) \hspace{-0.05in} - \hspace{-0.05in} \delta_3(l_1) \times
\bigg(
 1 - \sum_{k_U=1}^{K_\mathrm{U}} s_{m,k_U,i}[n] -  s^J_{m,i}[n]
\bigg) \bigg]^+  , \forall n, m, i, \label{eqn:varepsilon} \\
\hspace{-0.2in} \vartheta_m[n](l_1 \hspace{-0.05in} + \hspace{-0.05in} 1)\hspace{-0.15in} &=& \hspace{-0.15in} \bigg[ \vartheta_m[n](l_1) \hspace{-0.05in} - \hspace{-0.05in} \delta_4(l_1) \hspace{-0.05in} \times \hspace{-0.05in}
\bigg( \hspace{-0.05in}
P^m_\mathrm{peak} \hspace{-0.05in} - \hspace{-0.05in} \sum_{k_U=1}^{K_\mathrm{U}} \hspace{-0.05in} \sum_{i=1}^{N_\mathrm{F}} \widetilde{p}_{m,k_U,i}[n] \hspace{-0.05in} - \hspace{-0.05in} \sum_{i=1}^{N_\mathrm{F}} \hspace{-0.05in} s^J_{m,i}[n] p^J_{m,i}[n] \hspace{-0.05in}
\bigg) \bigg]^+, \forall n, m, \label{eqn:vartheta}
\end{eqnarray}
where $l_1 \geq 0$ is the iteration index for subproblem 1 and $\delta_u(l_1),u\in\{1,\ldots,4\}$, is the step size \cite{Ng_L}.
Thus, subproblem Layer 1 in problem \eqref{eqn:subproblem_1_dual problem} can be solved by using the updated Lagrangian multipliers in \eqref{eqn:alpha}-\eqref{eqn:vartheta}.
The proposed Algorithm for solving subproblem 1 is summarized in  {\bf Algorithm \ref{algorithm:algorithm 1}}.
Specifically, we solve the power allocation and user scheduling via the semi-closed-form solutions in \eqref{eqn:subproblem_1_optimal power} and \eqref{eqn:subproblem_1_optimal scheduling policy}, respectively, with a given Lagrange multipliers as shown in line 4 of {\bf Algorithm \ref{algorithm:algorithm 1}}.
Then, we update Lagrange multipliers via the gradient method \eqref{eqn:alpha}-\eqref{eqn:vartheta} as shown in line 6 of {\bf Algorithm \ref{algorithm:algorithm 1}}.

\begin{table}[t]\label{table:algorithm 1}\vspace*{-2mm}
  \begin{algorithm} [H]                    
  \renewcommand\thealgorithm{1}
  \caption{Optimal User Scheduling and Power Allocation for Subproblem 1}
  \label{algorithm:algorithm 1}     
   \begin{algorithmic} [1]
   \STATE Initialize the maximum number of iterations $L_{\rm{max\_inner}}^{l_1}$, $L_{\rm{max\_outer}}^{l_1}$, and the maximum tolerance $\epsilon^{l_1}_{\rm{inner}}$, $\epsilon^{l_1}_{\rm{outer}}$ for inner loop and outer loop, respectively.

   \STATE Set intermediate average minimum secrecy rate $\eta^{(0)} = 0$, iteration index $l_1^{\rm{inner}} = 0$, and $l_1 = 0$ for inner loop and outer loop, respectively.

   \REPEAT[Power Allocation and User Scheduling]

   \STATE  Maximize the Lagrangian over the minimum secrecy rate $\eta$, user scheduling $\mathcal{S}_U$, and power allocation $\tilde{\mathcal{P}}_U$ in \eqref{eqn:subproblem_1_dual problem} with the given Lagrange multipliers ${\bm\alpha},{\bm\beta}, {\bm\varepsilon},{\bm\vartheta}$.

   \REPEAT[Master Problem]

   \STATE Minimize the Lagrangian function in \eqref{eqn:subproblem_1_dual problem} over the Lagrange multipliers ${\bm\alpha} ,{\bm\beta}, {\bm\varepsilon}, {\bm\vartheta}$, for a fixed minimum secrecy rate $\eta$, user scheduling $\mathcal{S}_U$, and power allocation $\tilde{\mathcal{P}}_U$.

   \UNTIL{Convergence = \TRUE $\,$ or $l_1^{\rm{inner}} = L_{\rm{max\_inner}}^{l_1}$}.

   \UNTIL{Convergence = \TRUE $\,$ or $l_1 = L_{\rm{max\_outer}}^{l_1}$}.
   \STATE $\eta^{*l_1} = \eta^{l_1}$, $\mathcal{S}_U^* = \mathcal{S}_U^{l_1}$, and $\tilde{\mathcal{P}}_U^* = \tilde{\mathcal{P}}_U^{l_1}$.
   \end{algorithmic}
  \end{algorithm}\vspace*{-15mm}
\end{table}

\vspace*{-2mm}
\subsection{Subproblem 2: Jamming Policy}\vspace*{-1.5mm}

In this section, we consider subproblem 2 for optimizing jamming scheduling and power $\left(\mathcal{S}_J, \mathcal{P}_J\right)$ with the fixed communication policy and trajectories $(\mathcal{S}_U, \mathcal{P}_U,\mathcal{Q})$.
Thus, subproblem 2 can be written as\vspace*{-2mm}
\begin{align}\label{eqn:subproblem 2}
\underset{\eta,\, \mathcal{S}_J, \mathcal{P}_J}{\maxo}\,\,
&  \eta \\
  \mathrm{s.t.}\,
  \mathrm{C1}:\; &\frac{1}{N} \sum_{n=1}^{N}  \sum_{m=1}^{M} \sum_{i=1}^{N_\mathrm{F}}  R_{m,k_U,i}[n] - R'_{m,k_U,k_E,i}[n]  \geq \eta, \forall k_U,k_E, \notag
  \\
  \mathrm{C3}:\;  &s^J_{m,i}[n]\in \{0,1\}, \forall n, m, i,\notag
  \\
  \mathrm{C5}:\;  &\sum\limits_{{k_U} = 1}^{{K_\mathrm{U}}} s_{m,k_U,i}[n] +  s^J_{m,i}[n] \leq1, \forall n, m, i, \notag
  \\
  \mathrm{C6a}:\;  &  \sum\limits_{{k_U} = 1}^{{K_\mathrm{U}}} \sum_{i=1}^{N_\mathrm{F}}s_{m,k_U,i}[n] p_{m,k_U,i}[n] + \sum_{i=1}^{N_\mathrm{F}} s^J_{m,i}[n] p^J_{m,i}[n]\le P^m_\mathrm{peak},\; \forall n, m,\notag
  \\
  \mathrm{C6c}:\; &   p^J_{m,i}[n]   \ge 0, \forall n,m,i.\notag
\end{align}

First, we handle the coupling between binary variables $s^J_{m,i}[n]$ and continuous variables ${p}^J_{m,i}[n]$ in \eqref{eqn:subproblem 2}.
We introduce an auxiliary variable $\overline{p}^J_{m,i}[n] = s^J_{m,i}[n] p^J_{m,i}[n]$ and adopt the big-M formulation \cite{wei2016power} to transform problem \eqref{eqn:subproblem 2} equivalently as follows\vspace*{-2mm}
\begin{align}\label{eqn:sub2 s relax}
\underset{\eta,\, \mathcal{S}_J, \mathcal{P}_J, \overline{\mathcal{P}}_J}{\maxo}\,\,
&  \eta \\
  \mathrm{s.t.}\quad\; &\mathrm{C3}, \mathrm{C5}, \mathrm{C6c}, \notag
  \\
  &\overline{\mathrm{C1}}:\; \frac{1}{N} \sum_{n=1}^{N}  \sum_{m=1}^{M}  \sum_{i=1}^{N_\mathrm{F}}  \overline{R}_{m,k_U,i}[n] - \overline{R'}_{m,k_U,k_E,i}[n]  \geq \eta, \forall k_U,k_E, \notag
  \\
  &\overline{\mathrm{C6a}}:\; \sum\limits_{{k_U} = 1}^{{K_\mathrm{U}}} \sum_{i=1}^{N_\mathrm{F}} s_{m,k_U,i}[n] p_{m,k_U,i}[n] + \sum_{i=1}^{N_\mathrm{F}} \overline{p}^J_{m,i}[n]\le P^m_\mathrm{peak},\; \forall n, m,\notag
  \\
  &{\mathrm{C12}}:\;  \overline{p}^J_{m,i}[n] \ge 0, \forall n,m,i, \quad
   {\mathrm{C13}}:\;  \overline{p}^J_{m,i}[n] \le {p}^J_{m,i}[n], \forall n,m,i,    \notag
  \\
  &{\mathrm{C14}}:\;  \overline{p}^J_{m,i}[n] \le s^J_{m,i}[n] P^{Um}_\mathrm{peak}[n], \forall n,m,i,     \notag
  \\
  &{\mathrm{C15}}:\;  \overline{p}^J_{m,i}[n] \ge p^J_{m,i}[n] - (1-s^J_{m,i}[n])P^{Um}_\mathrm{peak}[n],     \forall n,m,i,\notag
\end{align}
where $\overline{\mathcal{P}}_J = \{ \overline{p}^J_{m,i}[n] , \forall  n, m, i \}$,\vspace*{-2mm}
\begin{align}\label{eqn:subproblem 2 more detail}
\overline{R}_{m,k_U,i}[n]
\hspace{-0.05in}  &=  \hspace{-0.05in}
s_{m,k_U,i}[n] \hspace{-0.03in} \log_2 \hspace{-0.03in} \left( \hspace{-0.03in} 1 + \frac{p_{m,k_U,i}[n]h_{m,k_U}[n]}{ \overline{I}_{m,k_U,i}[n] +\sigma^2} \right), \forall n, m, k_U, i,
\\
\overline{R}'\hspace{-0.04in}_{\hspace{-0.01in}m\hspace{-0.005in},\hspace{-0.01in}
k\hspace{-0.01in}_U\hspace{-0.005in},\hspace{-0.01in}k\hspace{-0.01in}_E
\hspace{-0.005in},\hspace{-0.01in}i}\hspace{-0.02in}[n]
\hspace{-0.05in}  &=  \hspace{-0.05in}
s_{m,k_U,i}[n] \hspace{-0.03in} \log_2 \hspace{-0.03in} \left( \hspace{-0.03in} 1 +  \frac{p_{m,k_U,i}[n]h_{m,k_E}[n]}{ \overline{I}_{m,k_E,i}[n] +\sigma^2} \right), \forall n, m, k_U, k_E, i,
\\
P^{Um}_\mathrm{peak}[n]
\hspace{-0.05in}  &=  \hspace{-0.05in}
P^m_\mathrm{peak} - \sum\limits_{{k_U} = 1}^{{K_\mathrm{U}}} \sum_{i=1}^{N_\mathrm{F}} s_{m,k_U,i}[n] p_{m,k_U,i}[n], \forall n, m,
\\
\overline{I}_{m,k_U,i}[n] &= \sum_{m'=1,m'\neq m}^{M} \overline{p}^J_{m',i}[n]h_{m',k_U}[n], \forall n, m, k_U,i,
\\
\overline{I}_{m,k_E,i}[n] &=  \sum_{m'=1,m'\neq m}^{M}  \overline{p}^J_{m',i}[n] h_{m',k_E}[n], \forall n, m, k_E,i,
\end{align}
and constraints ${\mathrm{C12}}$-${\mathrm{C15}}$ are imposed additionally due to the application of the big-M formulation \cite{wei2016power}.
Furthermore, the binary constraint $\mathrm{C3}$ is another major obstacle for designing a computationally-efficient jamming algorithm.
Therefore, the binary constraint $\mathrm{C3}$ can be written in its equivalent form:\vspace*{-4mm}
\begin{align}\label{eqn:subproblem 2 more detail}
\mathrm{C3a}: \; &   0 \le s^J_{m,i}[n] \le 1 \quad  \mathrm{and}   \notag \\
\mathrm{C3b}: \; &  \sum_{n=1}^{N}  \sum_{m=1}^{M} \sum_{i=1}^{N_\mathrm{F}} s^J_{m,i}[n] - \sum_{n=1}^{N}  \sum_{m=1}^{M} \sum_{i=1}^{N_\mathrm{F}} \left( s^{J}_{m,i}[n] \right)^2 \le 0.
\end{align}

However, the optimization problem is still hard to solve due to the non-convex constraints $\overline{\mathrm{C1}}$ and $\overline{\mathrm{C3b}}$.
Note that $\overline{\mathrm{C1}}$ and $\overline{\mathrm{C3b}}$ are in the form of the difference of two convex functions w.r.t. $\overline{p}^J_{m,i}[n]$ and ${s}^J_{m,i}[n]$, respectively \cite{a_li_cooperative_2019}.
Thus, following a similar approach as in \cite{wei2016power}, \cite{Optimal_Joint_Power_and_Subcarrier_Allocation}, we can augment the difference of convex (D.C.) constraints $\overline{\mathrm{C1}}$ and $\overline{\mathrm{C3b}}$ into the objective function with two penalty factors $\zeta$ and $\varphi$, which result in an equivalent problem.
Hence, problem \eqref{eqn:sub2 s relax} can be rewritten in its equivalent canonical form of D.C. programming as following\vspace*{-3mm}
\begin{align}\label{eqn:sub problem 2_DCD}
\underset{\eta,\, \mathcal{S}_J, \mathcal{P}_J, \overline{\mathcal{P}}_J}{\maxo}\,\, &F_1(\eta, { \mathcal{S}_J}, \overline{\mathcal{P}}_J  ) - F_2({ \mathcal{S}_J}, \overline{\mathcal{P}}_J)\\
\mathrm{s.t.}\quad\quad &{\mathrm{C3a}},{\mathrm{C5}}, {\overline{\mathrm{C6a}}},\mathrm{C6c},{\mathrm{C12}}-{\mathrm{C15}},\notag
\end{align}
\noindent where $F_1(\eta, \mathcal{S}_J, \overline{\mathcal{P}}_J )$ and $F_2(\mathcal{S}_J, \overline{\mathcal{P}}_J)$ are given by\vspace*{-1mm}
\begin{align}\label{eqn:subproblem 2 DCD F}
\hspace{-0.05in}F_1({\eta, \mathcal{S}_J, \overline{\mathcal{P}}_J}) =&
\eta \hspace{-0.03in} - \hspace{-0.03in} N K_U K_E \zeta  \eta \hspace{-0.03in} - \hspace{-0.03in}
\varphi \hspace{-0.03in} \sum\limits_{n = 1}^N  {\sum\limits_{m = 1}^M {\sum\limits_{i = 1}^{{N_\mathrm{F}}} \hspace{-0.03in} {s_{m,i}^J  \left[ n \right]} } }
\hspace{-0.03in} + \hspace{-0.03in}
\zeta \sum\limits_{\Upsilon} \hspace{-0.03in} \Big[ {s_{m,{k_U},i}}  \left[ n \right]
{{\log }_2}  \left( \overline{I}_{m,k_E,i}[n] \hspace{-0.03in} + \hspace{-0.03in} {\sigma ^2}  \right) \hspace{-0.03in} \Big]
\notag\\
&+  \zeta \sum\limits_{\Upsilon} \Big[
{s_{m,{k_U},i}}  \left[ n \right]  {{\log }_2}
\left( \overline{I}_{m,k_U,i}[n]  + {\sigma ^2}
 +
{p_{m,{k_U},i}\left[ n \right]}{h_{m,{k_U}}}  \left[ n \right] \right) \Big],
\\
\hspace{-0.05in}F_2({\mathcal{S}_J, \overline{\mathcal{P}}_J})  =&
\zeta  \sum\limits_{\Upsilon} \Big[
{{s_{m,{k_U},i}}  \left[ n \right]  {{\log }_2}
\left( \overline{I}_{m,k_E,i}[n]  +  {\sigma ^2}  +  {p_{m,{k_U},i} \left[ n \right]}{h_{m,{k_E}}}  \left[ n \right] \right)}  \Big]   \notag\\
&-
\varphi \sum\limits_{n = 1}^N  {\sum\limits_{m = 1}^M  {\sum\limits_{i = 1}^{{N_\mathrm{F}}}  {\left(s_{m,i}^{J}\right)^2  \left[ n \right]} } }
+
\zeta \sum\limits_{\Upsilon} \Big[   {{s_{m,{k_U},i}}   \left[ n \right]
{\log }_2\left( \overline{I}_{m,k_U,i}[n]  +  {\sigma ^2} \right)} \Big],
\end{align}
and we define $\sum\limits_{\Upsilon}[\cdot]$ as $\sum\limits_{n = 1}^N \sum\limits_{m = 1}^M \sum\limits_{{k_U=1}}^{{K_\mathrm{U}}}  \sum\limits_{{k_E} = 1}^{{K_\mathrm{E}}} \sum\limits_{i = 1}^{{N_\mathrm{F}}}[\cdot]$ for notational simplicity, where $\Upsilon \triangleq \{ \mathcal{N}, \mathcal{M}$, $\mathcal{K}_{U}, \mathcal{K}_{E}, \mathcal{N}_\mathrm{F} \}$.
Although it is still hard to solve the non-convex problem \eqref{eqn:sub problem 2_DCD} optimally, by utilizing the technique of SCA, we can obtain a locally optimal solution for problem \eqref{eqn:sub problem 2_DCD} \cite{huang2019cognitive}.
For the SCA technique, with a given feasible point at each iteration, the non-convex constraints can be approximated by the corresponding convex constraints, such that an approximated convex optimization problem can be obtained.
Then, by iteratively solving the sequence of approximated convex problems, an efficient solution to the original non-convex optimization problem \eqref{eqn:subproblem 2} can be obtained\cite{huang2019cognitive}.

Note that $F_1({\eta, \mathcal{S}_J, \overline{\mathcal{P}}_J})$ and $F_2({\mathcal{S}_J, \overline{\mathcal{P}}_J})$ are differentiable concave functions w.r.t. $\eta,\mathcal{S}_J$, and $\overline{\mathcal{P}}_J$.
Thus, for any feasible point $(\eta^{l_2},\mathcal{S}^{l_2}_J  ,\overline{\mathcal{P}}^{l_2}_J)$, we can define a global upper estimator for $F_2({\mathcal{S}_J, \overline{\mathcal{P}}_J})$  based on its first order Taylor's expansion at $(\eta^{l_2}, \mathcal{S}^{l_2}_J  ,\overline{\mathcal{P}}^{l_2}_J)$ as follows\vspace*{-2mm}
\begin{align}\label{eqn:sub2 inequality of DCD}
&F_2({{\mathcal{S}}_J \hspace{-0.01in}, \hspace{-0.01in} \overline{\mathcal{P}_J}})  \le  F_2(\mathcal{S}^{l_2}_J \hspace{-0.03in} , \hspace{-0.03in}\overline{\mathcal{P}}^{l_2}_J)
\hspace{-0.05in} + \hspace{-0.05in}
{\nabla _{\mathcal{S}_J}}F_2{(\mathcal{S}^{l_2}_J  \hspace{-0.03in} , \hspace{-0.03in} \overline{\mathcal{P}}^{l_2}_J)^T}({\mathcal{S}_J} \hspace{-0.05in} - \hspace{-0.05in} \mathcal{S}^{l_2}_J )
&\hspace{-0.15in} +
{\nabla _{\overline{\mathcal{P}}_J}} F_2{(\mathcal{S}^{l_2}_J  \hspace{-0.03in} , \hspace{-0.03in}\overline{\mathcal{P}}^{l_2}_J)^T} (\overline{\mathcal{P}}_J \hspace{-0.05in} - \hspace{-0.05in} \overline{\mathcal{P}}^{l_2}_J),
\end{align}
\noindent where ${\nabla\hspace{-0.08cm} _{\overline{\mathcal{P}}\hspace{-0.08cm}_J}} F_2{(\mathcal{S}^{l_2}_J  ,\hspace{-0.08cm}\overline{\mathcal{P}}^{l_2}_J)}$
and ${\nabla\hspace{-0.08cm} _{\mathcal{S}\hspace{-0.08cm}_J}}F_2{(\mathcal{S}^{l_2}_J  ,\hspace{-0.08cm}\overline{\mathcal{P}}^{l_2}_J)}$
denote the gradient vectors of $F_2({\mathcal{S}_J,\hspace{-0.08cm} \overline{\mathcal{P}}_J})$ at $(\mathcal{S}^{l_2}_J  ,\hspace{-0.08cm}\overline{\mathcal{P}}^{l_2}_J)$.

Moreover, the right hand side of \eqref{eqn:sub2 inequality of DCD} is an affine function.
Thus, we can obtain a lower bound for the optimal value of problem \eqref{eqn:sub problem 2_DCD} by solving the following concave maximization problem:\vspace*{-2mm}
\begin{align}\label{eqn:subproblem 2 final p}
&\underset{{\eta,\, \mathcal{S}_J, \mathcal{P}_J, \overline{\mathcal{P}}_J}}{\maxo}  F_1({\eta,\hspace{-0.03in} \mathcal{S}_J\hspace{-0.03in},\hspace{-0.03in} \overline{\mathcal{P}}_J})\hspace{-0.05in}
- \hspace{-0.05in}
F_2(\mathcal{S}^{l_2}_J\hspace{-0.03in},\hspace{-0.03in} \overline{\mathcal{P}}^{l_2}_J)
\hspace{-0.05in} - \hspace{-0.05in}
{\nabla\hspace{-0.03in} _{\mathcal{S}_J}}F_2{(\mathcal{S}^{l_2}_J\hspace{-0.03in},\hspace{-0.03in} \overline{\mathcal{P}}^{l_2}_J)^T}({\mathcal{S}_J}
\hspace{-0.05in} -  \hspace{-0.05in}
\mathcal{S}^{l_2}_J )
\hspace{-0.05in} - \hspace{-0.05in}
{\nabla\hspace{-0.03in} _{\overline{\mathcal{P}}_J}} F_2{(\mathcal{S}^{l_2}_J  \hspace{-0.03in},\hspace{-0.03in}\overline{\mathcal{P}}^{l_2}_J)^T} (\overline{\mathcal{P}}_J \hspace{-0.05in} -  \hspace{-0.05in} \overline{\mathcal{P}}^{l_2}_J)
\notag\\
&\quad\quad{\rm{s.t.}}\quad  {\mathrm{C3a}},{\mathrm{C5}}, {\overline{\mathrm{C6a}}},\mathrm{C6c},{\mathrm{C12}}-{\mathrm{C15}},
\end{align}
where
\begin{align}\label{eqn:subP2_de s}
\hspace*{-4mm}{\nabla _{\hspace*{-0.8mm}\mathcal{S}_{\hspace*{-0.8mm}J}}} \hspace*{-0.3mm} F_2 \hspace*{-0.3mm} {(\mathcal{S}^{l_2}_J  ,\overline{\mathcal{P}}^{l_2}_J)^{\hspace*{-0.3mm}T}}({\mathcal{S}_J} \hspace*{-1mm} -  \hspace*{-1mm} \mathcal{S}^{l_2}_J )
& \hspace*{-1mm} = \hspace*{-1mm}
-2\varphi  \sum\limits_{n = 1}^N\sum\limits_{m = 1}^M\sum\limits_{i = 1}^{N_\mathrm{F}} s^{J\left(l_2 \right)}_{m,i}[n] \left( s^J_{m,i}[n] \hspace*{-1mm} - \hspace*{-1mm} s^{J\left(l_2 \right)}_{m,i}[n]\right),
\\
\hspace*{-4mm}{\nabla _{\hspace*{-0.8mm}\overline{\mathcal{P}}_{\hspace*{-0.8mm}J}}} \hspace*{-0.3mm} F_2 \hspace*{-0.3mm}{(\mathcal{S}^{l_2}_J  ,\overline{\mathcal{P}}^{l_2}_{J})^{\hspace*{-0.3mm}T}} \hspace*{-0.3mm} (\overline{\mathcal{P}}_{\hspace*{-1mm}J}
\hspace*{-1mm} -  \hspace*{-1mm} \overline{\mathcal{P}}^{l_2}_{\hspace*{-1mm}J} )
%
%
& \hspace*{-1mm} = \hspace*{-1mm} \zeta \hspace*{-0.8mm} \sum\limits_{\Upsilon} \hspace*{-0.8mm} \Big[ \hspace*{-0.5mm}
\frac{s_{m,k_U,i}[n]}{\ln 2} \hspace*{-1mm}
\left( \hspace*{-0.3mm} A^{l_2}_{m,k_U,i}[n] \hspace*{-1mm} + \hspace*{-1mm}  B^{l_2}_{m,k_U,k_E,i}[n] \hspace*{-0.3mm} \right) \hspace*{-1.3mm}
\left( \overline{p}^J_{m,i}[n] \hspace*{-1mm} - \hspace*{-1mm} \overline{p}^{J\left( l_2 \right)}_{m,i}[n] \right) \hspace*{-1mm} \Big]  \hspace*{-0.5mm},
\\
A^{l_2}_{m,k_U,i}[n]
& \hspace*{-1mm}=
\frac{\sum\limits_{m' = 1, m'\neq m}^M h_{m',k_U}[n]}{\overline{I}^{l_2}_{m,k_U,i}[n] + \sigma^2}, \\
B^{l_2}_{m,k_U,k_E,i}[n]
&\hspace*{-1mm}=
\frac{\sum\limits_{m' = 1, m'\neq m}^M h_{m',k_E}[n]}{\overline{I}^{l_2}_{m,k_E,i}[n] + \sigma^2 + p_{m,k_U,i}[n]h_{m,k_E}[n]}.
\end{align}\vspace*{-2mm}

Now, the optimization problem in \eqref{eqn:subproblem 2 final p} is a convex optimization problem which can be solved efficiently by standard convex problem solvers, such as CVX \cite{Boyd}.
To tighten the obtained lower bound, we adopt an iterative algorithm to generate a sequence of feasible solutions successively, cf. {\bf Algorithm \ref{algorithm:algorithm 2}}.
The initial feasible solution with iteration index $l_2=0$ is obtained by solving the convex optimization problem in \eqref{eqn:subproblem 2 final p} with $F_1({\eta, \mathcal{S}_J, \overline{\mathcal{P}}_J})$ as the objective function \cite{wei2016power} which is shown in line 2 of {\bf Algorithm \ref{algorithm:algorithm 2}}.
Then, the intermediate solution from the last iteration will be used to update the problem in \eqref{eqn:subproblem 2 final p} and it will generate a feasible solution for the next iteration in $l_2 = l_2 +1$, as shown in line 5 of {\bf Algorithm \ref{algorithm:algorithm 2}}.
The iterative procedure will stop either the changes of optimization variables are smaller than a predefined convergence tolerance or the number of iteration reaches its maximum.\vspace*{-1.5mm}

\begin{table}[t]\label{table:algorithm 2}\vspace*{-2mm}
  \begin{algorithm} [H]                
  \renewcommand\thealgorithm{2}
  \caption{Jamming Policy Optimization Algorithm}
  \label{algorithm:algorithm 2}
   \begin{algorithmic} [1]
  \STATE Initialize the maximum number of iterations $L_{\max}^{l_2}$, iteration index $l_2=0$, penalty factors $\zeta$ and $\varphi$, and the maximum tolerance $\epsilon^{l_2}$.

  \STATE  Set the intermediate average minimum secrecy rate, jamming scheduling, jamming power, and relaxed jamming power as $\eta^{(0)}$, $\mathcal{S}_J^{(0)}$, $\mathcal{P}_J^{(0)}$, and $\overline{\mathcal{P}}_J^{(0)}$, respectively.

  \REPEAT
  \STATE Solve \eqref{eqn:subproblem 2 final p} for a given communication resource allocation $\{\mathcal{S}_C, \mathcal{P}_C\}$ and UAVs' trajectories $\{\mathcal{Q}\}$.
  \STATE Set $l_2 = l_2+1$, $\eta^{l_2} = \eta$, $\mathcal{S}_J^{l_2}=\mathcal{S}_J$, $\mathcal{P}_J^{l_2}=\mathcal{P}_J$, and $\overline{\mathcal{P}}_J^{l_2}=\overline{\mathcal{P}}_J$.

  \UNTIL convergence or $l_2=L_{\max}^{l_2}$.

  \STATE $\eta^{l_2} = \eta^{l_2}$, $\mathcal{S}_J= \mathcal{S}_J^{l_2}$, $\mathcal{P}_J = \mathcal{P}_J^{l_2}$, and $\overline{\mathcal{P}}_J = \bold \overline{\mathcal{P}}_J^{l_2}$.

  \end{algorithmic}
  \end{algorithm}
  \vspace*{-15mm}
\end{table}

\vspace*{-3mm}
\subsection{Subproblem 3: Trajectory Optimization}\vspace*{-1.5mm}

In this section, we consider subproblem 3 for optimizing the trajectory design by assuming that $(\mathcal{S}_U, \mathcal{S}_J, \mathcal{P}_U, \mathcal{P}_J)$ are fixed.
Thus, subproblem 3 can be written as\vspace*{-2mm}
\begin{align}\label{eqn:subproblem 3}
\underset{\eta,\,\mathcal{Q}}{\maxo}\,\,
&  \eta \\
  \mathrm{s.t.}\,
  \mathrm{C1}:\; &\frac{1}{N} \sum_{n=1}^{N}  \sum_{m=1}^{M}  \sum_{i=1}^{N_\mathrm{F}}   R_{m,k_U,i}[n] - R'_{m,k_U,k_E,i}[n]  \geq \eta, \forall k_U,k_E, \notag
  \\
   \mathrm{C7}:\;  &|| \mathbf{q}_m[n] - \mathbf{q}_m\left[n-1\right] ||^2 \leq V^2, \forall n,m, \notag\\
  \mathrm{C8}:\;  &|| \mathbf{q}_m[n] - \mathbf{w}_\mathrm{NF}^j ||^2  \geq \left( Q_\mathrm{NF}^{j} \right)^2, \forall n, m, j, \notag
  \\
  \mathrm{C9}:\;  &|| \mathbf{q}_m[n] - \mathbf{q}_{m'}\left[n\right] ||^2 \geq \hspace*{-1mm} D_\mathrm{S}^2, \forall n,m, m \hspace*{-1mm} \neq \hspace*{-1mm} m',  \notag
  \\
  \mathrm{C10}:\;  &\mathbf{q}_{m}[0] = \mathbf{q}_{m}^0, \forall m, \quad\quad\quad\quad\quad\quad\;
  \mathrm{C11}:\;  \mathbf{q}_{m}[N]  = \mathbf{q}_{m}^F, \forall m.\notag
\end{align}

However, the optimization problem is still non-convex due to constraints $\mathrm{C1}$, $\mathrm{C8}$, and $\mathrm{C9}$.
To facilitate the development of solution, we first introduce four slack variables $\bold{t}_{U} = \{ t_{m,k_U}[n], \forall n,m,k_U \}$, $\bold{t}'_{U} = \{ t'_{m,k_U}[n], \forall n,m,k_U \}$, $\bold{t}_{E} = \{ t_{m,k_E}[n], \forall n,m,k_E \}$, and $\bold{t}'_{E} = \{ t'_{m,k_E}[n], \forall n,m,k_E \}$, which satisfy\vspace*{-2mm}
\begin{align}\label{eqn:sub problem 3 slack t}
&\mathrm{C12}:\;\;  t_{m,k_U}[n] \ge d^2_{m,k_U}[n], \forall n,m,k_U,
\quad
\mathrm{C13}:\;\;  t_{m,k_E}[n] \ge d^2_{m,k_E}[n], \forall n,m,k_E,
\\
&\mathrm{C14}:\;\;  t'_{m,k_U}[n] \le d^2_{m,k_U}[n], \forall n,m,k_U,
\quad
\mathrm{C15}:\;\;  t'_{m,k_E}[n] \le d^2_{m,k_E}[n], \forall n,m,k_E.
\end{align}

Then, communication rate $R_{m,k_U,i}[n]$  in constraint $\mathrm{C1}$ can be written as
%
\begin{align}\label{eqn:sub problem 3 com rate}
{R}_{m,k_U,i}[n] \hspace*{-0.6mm}
  = \hspace*{-0.6mm} s_{m,k_U,i}[n] \log_2\left(1 + \frac{ \frac{p_{m,k_U,i}[n]\beta_0}{t_{m,k_U}[n]} } {\Omega_{m,k_U}[n]} \right) \hspace*{-0.6mm}
  =\hspace*{-0.6mm} \hat{R}_{m,k_U,i}[n] \hspace*{-0.6mm} - \hspace*{-0.6mm}  \check{R}_{m,k_U,i}[n],\forall n,m,k_U,i,
\end{align}\vspace*{-2mm}
where\vspace*{-2mm}
\begin{align}\label{eqn:sub problem 3 com rate R1 and R2}
\hat{R}_{m,k_U,i}[n]
  &= s_{m,k_U,i}[n] \log_2 \left(\frac{p_{m,k_U,i}[n]\beta_0}{t_{m,k_U}[n]}
   +  \Omega_{m,k_U}[n]  \right),
\\
\check{R}_{m,k_U,i}[n] &=
  s_{m,k_U,i}[n] \log_2 \Omega_{m,k_U}[n],
\\
\Omega_{m,k_U}[n] &= { \sum\limits_{m'=1,m'\neq m}^M   \frac{s^J_{m',i}[n] p^J_{m',i}[n]\beta_0}
  { {t'_{m,k_U}[n]} } +\sigma^2}.
\end{align}

Similarly, the leakage rate in constraint $\mathrm{C1}$ can be written as
%
\begin{align}\label{eqn:sub problem 3 leakage rate}
\hspace*{-1.5mm}{R}'_{\hspace*{-0.1mm}m\hspace*{-0.2mm},\hspace*{-0.2mm}k_U\hspace*{-0.2mm},
\hspace*{-0.2mm}k_E\hspace*{-0.2mm},\hspace*{-0.2mm}i}\hspace*{-0.2mm}[n] \hspace*{-1.1mm}
=  \hspace*{-1.1mm} s_{m,k_U,i}[n]\hspace*{-1.1mm}
\log_2 \hspace*{-1.2mm} \left( \hspace*{-1.5mm} 1 \hspace*{-1.2mm}  +  \hspace*{-1.2mm} \frac{ \frac{ p_{m,k_U,i}[n]\beta_0} {t_{m,k_E}[n]}}{\Omega_{m,k_E}[n]} \right)
\hspace*{-1.5mm} = \hspace*{-1.5mm}
\check{R}'_{m,k_U,k_E,i}[n] \hspace*{-1.5mm} - \hspace*{-1.5mm} \hat{R}'_{m,k_U,k_E,i}[n], \forall n\hspace*{-0.5mm},\hspace*{-0.5mm}m\hspace*{-0.5mm},\hspace*{-0.5mm}k_U\hspace*{-0.5mm},
\hspace*{-0.5mm}k_E\hspace*{-0.5mm},\hspace*{-0.5mm}i,\hspace*{-0.5mm}
\end{align}\vspace*{-2mm}
where\vspace*{-2mm}
\begin{align}\label{eqn:sub problem 3 leakage rate R3 and R4}
\check{R}'_{m,k_U,k_E,i}[n]
  & =    s_{\hspace*{-0.1mm}m\hspace*{-0.1mm},\hspace*{-0.1mm}k_U\hspace*{-0.1mm},\hspace*{-0.1mm}i}\hspace*{-0.1mm}[n]
 \hspace*{-0.6mm} \log_2 \hspace*{-0.6mm} \left( \frac{ p_{m,k_U,i}[n]\beta_0} {t_{m,k_E}[n] }
  \hspace*{-0.6mm} + \hspace*{-0.6mm}
  \Omega_{m,k_E}[n] \right),
\\
\hat{R}'_{m,k_U,k_E,i}[n]  & =
 s_{\hspace*{-0.1mm}m\hspace*{-0.1mm},\hspace*{-0.1mm}k_U\hspace*{-0.1mm},\hspace*{-0.1mm}i}\hspace*{-0.1mm}[n]
  \hspace*{-0.6mm} \log_2 \hspace*{-0.6mm} \Omega_{m,k_E}[n],
  \\
\Omega_{m,k_E}[n] &= { \sum\limits_{m'=1,m'\neq m}^M \frac{ s^J_{m',i}[n] p^J_{m',i}[n] \beta_0 } {t'_{m,k_E}[n]} +\sigma^2}.
\end{align}

Therefore, the problem can be written as\vspace*{-2mm}
\begin{align}\label{eqn:sub problem 3 new C1}
&\underset{\eta,\mathcal{Q},\bold{t}_U,\bold{t}_E,\bold{t}'_U,\bold{t}'_E }{\maxo}\,\,
  \eta \\
&\;\mathrm{s.t.}\quad   \mathrm{C7} - \mathrm{C11}, \notag\\
&\hat{\mathrm{C1}}:\;\; \frac{1}{N} \sum_{n=1}^{N}  \sum_{m=1}^{M}  \sum_{i=1}^{N_\mathrm{F}}   \hat{R}_{m,k_U,i}[n] + \hat{R}'_{m,k_U,k_E,i}[n] - \check{R}_{m,k_U,i}[n] - \check{R}'_{m,k_U,k_E,i}[n] \geq \eta, \forall k_U,k_E,
\notag \\
&\mathrm{C12}:\;\;  t_{m,k_U}[n] \ge d^2_{m,k_U}[n], \forall n,m,k_U,
\quad
\mathrm{C13}:\;\;  t_{m,k_E}[n] \ge d^2_{m,k_E}[n], \forall n,m,k_E,
\notag\\
&\mathrm{C14}:\;\;  t'_{m,k_U}[n] \le d^2_{m,k_U}[n], \forall n,m,k_U,
\quad
\mathrm{C15}:\;\;  t'_{m,k_E}[n] \le d^2_{m,k_E}[n], \forall n,m,k_E,
\notag
\end{align}
where $\bold{t}_U = \{t_{m,k_U}[n] , \forall n, m, k_U\}$, $\bold{t}_E = \{ t_{m,k_E}[n] , \forall n, m, k_E\}$, $\bold{t}'_U = \{t'_{m,k_U}[n] , \forall n, m, k_U\}$, and $\bold{t}'_E = \{ t'_{m,k_E}[n] , \forall n, m, k_E\}$.

Since $\hat{R}_{m,k_U,i}[n]$, $\check{R}_{m,k_U,i}[n]$, $\hat{R}'_{m,k_U,k_E,i}[n]$, and $\check{R}'_{m,k_U,k_E,i}[n]$ are convex functions w.r.t. $t_{m,k_U}[n]$, $t'_{m,k_U}[n]$, $t_{m,k_E}[n]$, and $t'_{m,k_E}[n]$, respectively, problem \eqref{eqn:sub problem 3 new C1} is equivalent to problem \eqref{eqn:subproblem 3}, as constraints $\mathrm{C12}$ - $\mathrm{C15}$ hold with equalities at the optimal point of problem  \eqref{eqn:sub problem 3 new C1}\cite{li,Zhong_Cooperative_Jamming}.

However, problem \eqref{eqn:sub problem 3 new C1} is still non-convex due to the non-convex constraints $\hat{\mathrm{C1}}$, $\mathrm{C8}$, $\mathrm{C9}$, $\mathrm{C14}$, $\mathrm{C15}$, $\mathrm{C16}$, and $\mathrm{C18}$.
Although it is hard to solve the non-convex problem \eqref{eqn:sub problem 3 new C1} optimally, similar to the case for solving subproblem 2, by utilizing the technique of SCA, we can obtain a locally optimal solution for problem \eqref{eqn:sub problem 3 new C1}.
To this end, we first handle constraint $\hat{\mathrm{C1}}$. Since $ - \check{R}_{m,k_U,i}[n]$ is a convex functions w.r.t. $\{t'_{m,k_U}[n]\}$.
We have the following inequalities by applying the first order Taylor expansion at any given point $\{t'_{m,k_U}[n]\}$\vspace*{-2mm}
\begin{align}\label{eqn:sub problem 3 SCA R2}
- \check{R}_{m,k_U,i}[n] &=
- s_{m,k_U,i}[n] \log_2 \Omega_{m,k_U}[n]
\notag \\
&\ge - \check{R}^{l_3}_{m,k_U,i}[n] - \nabla_{{t}'_U} \check{R}^{l_3}_{m,k_U,i}[n]
\left( { t'_{m,k_U}[n] - {t'}^{l_3}_{m,k_U}[n] } \right)
\notag \\
&\triangleq - \check{R}^{\rm{lb}}_{m,k_U,i}[n], \forall n,m,k_U,i,
\end{align}
where\vspace*{-4mm}
\begin{align}\label{eqn:sub problem 3 SCA R2 detail}
\check{R}^{l_3}_{m,k_U,i}[n]
& = s_{m,k_U,i}[n] \log_2 \Omega^{l_3}_{m,k_U}[n], \\
\nabla_{{t}'_U} \check{R}^{l_3}_{m,k_U,i}[n]
& = - \sum\limits_{m'=1,m'\neq m}^M
\frac{s_{m,k_U,i}[n] s^J_{m',i}[n] p^J_{m',i}[n]\beta_0  }
{\left( {t'}^{l_3}_{m',k_U}[n] \right)^2 \Omega^{l_3}_{m,k_U}[n] \ln 2}.
\end{align}

Similarly, we approximate $- \check{R}'_{m,k_U,k_E,i}[n]$ by applying the first order Taylor expansion at given points $\{t_{m,k_E}[n]\}$ and $\{t'_{m,k_E}[n]\}$\vspace*{-2mm}
\begin{align}\label{eqn:sub problem 3 SCA R3}
&\quad -\hspace*{-0.8mm} \check{R}'_{m,k_U,k_E,i}
\hspace*{-0.2mm}[\hspace*{-0.2mm}n\hspace*{-0.2mm}]\hspace*{-1.5mm}
=\hspace*{-0.8mm}
-s_{m,k_U,i}[n]
\hspace*{-0.3mm} \log_2 \hspace*{-0.3mm} \left( \frac{ p_{m,k_U,i}[n]\beta_0} {t_{m,k_E}[n] }
\hspace*{-0.3mm} + \hspace*{-0.3mm}
\Omega_{m,k_E}^{l_3}[n]  \right)
\notag \\
&\ge
\hspace*{-1mm} - \hspace*{-1mm}
\check{R}'^{l_3}_{m,k_U,k_E,i}
\hspace*{-0.2mm}[\hspace*{-0.2mm}n\hspace*{-0.2mm}]
\hspace*{-1mm} - \hspace*{-1.5mm}
\nabla_{\hspace*{-0.4mm}{t}_{\hspace*{-0.2mm} E}}\hspace*{-0.2mm} \check{R}'^{l_3}_{m,k_U,k_E,i}
\hspace*{-0.2mm}[\hspace*{-0.2mm}n\hspace*{-0.2mm}] \hspace*{-1mm}
\left( \hspace*{-0.4mm} { t_{m,k_E}\hspace*{-0.2mm}[\hspace*{-0.2mm}n\hspace*{-0.2mm}]
\hspace*{-1mm} - \hspace*{-1mm}
{t}^{l_3}_{m,k_E}\hspace*{-0.2mm}[\hspace*{-0.2mm}n\hspace*{-0.2mm}] } \hspace*{-0.4mm} \right)
\hspace*{-1mm} - \hspace*{-1.5mm}
\nabla_{\hspace*{-0.4mm}{t'}_{\hspace*{-0.2mm} E}}\hspace*{-0.2mm}  \check{R}'^{l_3}_{m,k_U,k_E,i}\hspace*{-0.2mm}[\hspace*{-0.2mm}n\hspace*{-0.2mm}] \hspace*{-1mm}
\left(\hspace*{-0.4mm} { t'_{m,k_E}\hspace*{-0.2mm}[\hspace*{-0.2mm}n\hspace*{-0.2mm}]
\hspace*{-1mm} - \hspace*{-1mm}
{t'}^{l_3}_{m,k_E}\hspace*{-0.2mm}[\hspace*{-0.2mm}n\hspace*{-0.2mm}] } \hspace*{-0.4mm}\right)
\notag \\
&\triangleq
\hspace*{-1mm} - \hspace*{-1mm}
\check{R}^{\rm{lb}}_{m,k_U,k_E,i}
\hspace*{-0.2mm}[\hspace*{-0.2mm}n\hspace*{-0.2mm}], \forall n,m,k_U,k_E,i,
\end{align}\vspace*{-2mm}
where\vspace*{-3mm}
\begin{align}\label{eqn:sub problem 3 SCA R3 detail}
\check{R}^{l_3}_{m,k_U,k_E,i}[n] &= s_{m,k_U,i}[n]
\log_2  \left( \frac{ p_{m,k_U,i}[n]\beta_0} {t^{l_3}_{m,k_E}[n] }
+ \Omega_{m,k_E}^{l_3}[n]  \right),
\\
%
\nabla_{{t}_{E}} \check{R}'^{l_3}_{m,k_U,k_E,i}[n] &=
\frac{ - s_{m,k_U,i}[n] p_{m,k_U,i}[n]\beta_0  } {\left( {t}^{l_3}_{m,k_E}[n] \right)^2
\left( \frac{ p_{m,k_U,i}[n]\beta_0} {t^{l_3}_{m,k_E}[n] }
+ \Omega_{m,k_E}^{l_3}[n] \right)  \ln  2 },
\\
\nabla_{{t'}_{E}} \check{R}'^{l_3}_{m,k_U,k_E,i}[n]
&= \sum\limits_{m'=1,m'\neq m}^M
\frac{ - s_{m,k_U,i}[n] s^J_{m',i}[n] p^J_{m',i}[n]\beta_0  }
{\left( {t'}^{l_3}_{m',k_E}[n] \right)^2
\left( \frac{ p_{m,k_U,i}[n]\beta_0} {t^{l_3}_{m,k_E}[n] }
+ \Omega_{m,k_E}^{l_3}[n]  \right)  \ln  2 }.
\end{align}

By replacing $ - \check{R}_{m,k_U,i}[n] - \check{R}'_{m,k_U,k_E,i}[n]$ in $\hat{\mathrm{C1}}$, constraint $\hat{\mathrm{C1}}$ can be written as\vspace*{-2mm}
\begin{align}\label{eqn:sub problem 3 new C1 SCA}
\hat{\hat{\mathrm{C1}}}:
\frac{1}{N} \sum_{n=1}^{N}  \sum_{m=1}^{M} \sum_{i=1}^{N_\mathrm{F}}   \hat{R}_{m,k_U,i}[n]
\hspace*{-1.2mm} + \hspace*{-1.2mm} \hat{R}'_{m,k_U,k_E,i}[n]
\hspace*{-1.2mm} - \hspace*{-1.2mm}
\check{R}^{\rm{lb}}_{\hspace*{-0.2mm}m,\hspace*{-0.2mm}k_U,\hspace*{-0.2mm}i}
\hspace*{-0.4mm}[\hspace*{-0.4mm}n\hspace*{-0.4mm}]
\hspace*{-1.2mm} - \hspace*{-1.2mm}
\check{R}^{\rm{lb}}_{\hspace*{-0.2mm}m,\hspace*{-0.2mm}k_U,\hspace*{-0.2mm}k_E,\hspace*{-0.2mm}i}
\hspace*{-0.4mm}[\hspace*{-0.4mm}n\hspace*{-0.4mm}]
 \geq \eta, \forall k_U,k_E.
\end{align}

Then, we handle the non-convex NFZ constraint $\mathrm{C8}$\vspace*{-2mm}
\begin{align}\label{eqn:sub problem 3 C8 SCA}
\hat{\mathrm{C8}}:
|| \mathbf{q}^{l_3}_m[n]  -  \mathbf{w}_\mathrm{NF}^j ||^2
+ 2\left(   \mathbf{q}^{l_3}_m[n]   -  \mathbf{w}_\mathrm{NF}^j \right)
\left(\mathbf{q}_m[n]  - \mathbf{q}^{l_3}_m[n] \right)
\geq \left( Q_{\mathrm{NF}}^{j}  \right)^{2}, \forall n,m,j.
\end{align}

Similarly, we approximate the non-convex constraint $\mathrm{C9}$, $\mathrm{C14}$, and $\mathrm{C15}$ as following\vspace*{-2mm}
\begin{align}\label{eqn:sub problem 3 C9 SCA}
\hat{\mathrm{C9}}:
|| \mathbf{q}^{l_3}_m[n] \hspace*{-1mm} - \hspace*{-1mm} \mathbf{q}^{l_3}_{m'}[n] ||^2
\hspace*{-1mm} + \hspace*{-1mm}
2\left(   \mathbf{q}^{l_3}_m[n]  \hspace*{-1mm} - \hspace*{-1mm} \mathbf{q}^{ l_3}_{m'}[n] \right) \hspace*{-1mm}
\left(\mathbf{q}_m[n] \hspace*{-1mm} - \hspace*{-1mm} \mathbf{q}^{l_3}_m[n] \right)
\geq D_{\mathrm{S}}^{2},
\forall n,m, m' \neq  m,
\end{align}
\begin{align}\label{eqn:sub problem 3 C14 SCA}
\hat{\mathrm{C14}}:
|| \mathbf{q}^{l_3}_m[n] \hspace*{-1mm} - \hspace*{-1mm} \mathbf{w}_{k_U} ||^2
\hspace*{-1mm} + \hspace*{-1mm}
2\left(   \mathbf{q}^{l_3}_m[n]  \hspace*{-1mm} - \hspace*{-1mm}  \mathbf{w}_{k_U} \right) \hspace*{-1mm}
\left(\mathbf{q}_m[n] \hspace*{-1mm} - \hspace*{-1mm} \mathbf{q}^{l_3}_m[n] \right)
+H^2
\ge  t'_{m,k_{U}}[n],
\forall  n,m,k_{U},
\end{align}
\begin{align}\label{eqn:sub problem 3 C15 SCA}
\hat{\mathrm{C15}}:
|| \mathbf{q}^{l_3}_m[n] \hspace*{-1mm} - \hspace*{-1mm} \mathbf{w}_{k_E} ||^2
\hspace*{-1mm} + \hspace*{-1mm}
2\left(   \mathbf{q}^{l_3}_m[n]  \hspace*{-1mm} - \hspace*{-1mm}  \mathbf{w}_{k_E} \right) \hspace*{-1mm}
\left(\mathbf{q}_m[n] \hspace*{-1mm} - \hspace*{-1mm} \mathbf{q}^{l_3}_m[n] \right)
+H^2
\ge  t'_{m,k_{E}}[n],  \forall  n,m,k_{E}.
\end{align}

Now, with the more stringent constraints $\hat{\hat{\mathrm{C1}}}$, $\hat{\mathrm{C8}}$, $\hat{\mathrm{C9}}$, $\hat{\mathrm{C14}}$, and $\hat{\mathrm{C15}}$, a suboptimal solution of \eqref{eqn:subproblem 3} can be obtained by solving the following optimization problem\vspace*{-2mm}
\begin{align}\label{eqn:sub problem 3 all convex}
&\underset{\eta, \mathcal{Q},\bold{t}_U,\bold{t}_E,\bold{t}'_U,\bold{t}'_E }{\maxo}\,\,
  \eta \\
&\;\mathrm{s.t.}\quad   \hat{\hat{\mathrm{C1}}}, \mathrm{C7}, \hat{\mathrm{C8}}, \hat{\mathrm{C9}}, \mathrm{C10} - \mathrm{C13}, \hat{\mathrm{C14}}, \hat{\mathrm{C15}},
\notag
\end{align}
which is a convex optimization problem and can be solved efficiently by standard convex problem solvers such as CVX \cite{Boyd}, which is summarized in {\bf Algorithm \ref{algorithm:algorithm 3}}.

\vspace*{-3mm}
\subsection{Overall Algorithm}\vspace*{-2.5mm}
In summary, the proposed algorithm solves the three subproblems \eqref{eqn:subproblem 1}, \eqref{eqn:subproblem 2}, and \eqref{eqn:subproblem 3} in an alternating manner with a polynomial time computational complexity.
The details of the proposed algorithm are summarized in {\bf Algorithm \ref{algorithm:algorithm 4}}.
Since the objective value of \eqref{eqn:problem 1} with the solutions obtained by solving subproblems \eqref{eqn:subproblem 1}, \eqref{eqn:subproblem 2}, and \eqref{eqn:subproblem 3} is non-decreasing over iterations and feasible solution set is bounded, the solution obtained by the proposed algorithm is guaranteed to converge to a suboptimal solution with a polynomial time computational complexity \cite{Zeng_2016_Throughput_Maximization_for_UAV}, \cite{wei2016power}.
\vspace*{-5mm}

\begin{table}[t]\label{table:algorithm 3}\vspace*{-8mm}
  \begin{algorithm} [H]                
  \renewcommand\thealgorithm{3}
  \caption{Trajectory Design Optimization Algorithm}
  \label{algorithm:algorithm 3}
   \begin{algorithmic} [1]
  \STATE Initialize the maximum number of iterations $L_{\max}^{l_3}$, iteration index $l^{l_3}=0$, and the maximum tolerance $\epsilon^{l_3}$.

  \STATE  Set the initial values as $\eta^{(0)}$, $\mathcal{Q}^{(0)}$, $\bold{t}_U^{(0)}$, $\bold{t}_E^{(0)}$, $\bold{{t}'}_U^{(0)}$, and $\bold{{t}'}_E^{(0)}$, respectively.\vspace*{-0.5mm}

  \REPEAT\vspace*{-0.5mm}
  \STATE Solve \eqref{eqn:sub problem 3 all convex} for a given communication resource allocation $\{\mathcal{S}_C, \mathcal{P}_C\}$ and jammer policy $\{\mathcal{S}_J, \mathcal{P}_J\}$.
  \STATE Set $l_3 = l_3+1$, $\eta^{l_3} = \eta$, $\mathcal{Q}^{l_3} = \mathcal{Q}$, $\bold{t}_U^{l_3} = \bold{t}_U$, $\bold{t}_E^{l_3} = \bold{t}_E$, $\bold{{t}'}_U^{l_3} = \bold{{t}'}_U$, and $\bold{{t}'}_E^{l_3} = \bold{{t}'}_E$.

  \UNTIL convergence or $l_3=L_{\max}^{l_3}$.

  \STATE $\eta^{l_3} = \eta^{l_3}$, $\mathcal{Q} = \mathcal{Q}^{l_3}$, $\bold{t}_U = \bold{t}_U^{l_3}$, $\bold{t}_E = \bold{t}_E^{l_3}$, $\bold{{t}'}_U = \bold{{t}'}_U^{l_3}$, and $\bold{{t}'}_E = \bold{{t}'}_E^{l_3}$.

  \end{algorithmic}
  \end{algorithm}
  \vspace*{-17mm}
\end{table}


\begin{table}[t]\label{table:algorithm 4}\vspace*{-2mm}
  \begin{algorithm} [H]
  \renewcommand\thealgorithm{4}
  \caption{Iterative Resource Allocation and Trajectory Optimization Algorithm}

  \label{algorithm:algorithm 4}
  \begin{algorithmic} [1]
  \STATE Initialize the maximum number of iterations $L_{\rm{max}}^{l_4}$, iteration index $l_4=0$, and the maximum tolerance $\epsilon^{l_4}$.\vspace*{-0.5mm}

  \REPEAT \vspace*{-0.5mm}

  \STATE For the fixed jamming policy and trajectory, obtain the optimal communication resource allocation $\{\mathcal{S}_U^*, \mathcal{P}_U^*\}$ and intermediate average minimum secrecy rate $\eta^{l_1}$ by using Algorithm 1.

  \STATE For the fixed communication resource allocation and trajectory, obtain the jamming policy $\{\mathcal{S}_J, \mathcal{P}_J\}$ and intermediate average minimum secrecy rate $\eta^{l_2}$ by using Algorithm 2.

  \STATE For the fixed communication resource allocation and jamming policy, obtain the trajectory $\mathcal{Q}$ and intermediate average minimum secrecy rate $\eta^{l_3}$ by using Algorithm 3.

  \STATE Set $l_4 = l_4+1$,
  $ \eta^{l_4} = \eta^{l_3}$,
  $\{\mathcal{S}_U^{l_4}, \mathcal{P}_U^{l_4} \} = \{\mathcal{S}_U^*, \mathcal{P}_U^*\}$,
  $\{\mathcal{S}_J^{l_4}, \mathcal{P}_J^{l_4}\} = \{\mathcal{S}_J, \mathcal{P}_J\}$,
  and $\mathcal{Q}^{l_4}= \mathcal{Q}$.

  \UNTIL convergence or iteration index reaches to the maximum number.

  \STATE $\eta = \eta^{l_4}$, $\{\mathcal{S}_U, \mathcal{P}_U\} = \{\mathcal{S}_U^{l_4}, \mathcal{P}_U^{l_4}\}$, $\{\mathcal{S}_J, \mathcal{P}_J\} = \{\mathcal{S}_J^{l_4}, \mathcal{P}_J^{l_4}\}$, and $\mathcal{Q} = \mathcal{Q}^{l_4}$.

  \end{algorithmic}
  \end{algorithm}
  \vspace*{-17.5mm}
\end{table}

\section{Numerical Results}\vspace*{-2.5mm}

In this section, we investigate the performance of the proposed UAV-enabled secure communication scheme through simulations.
Unless specified otherwise, the system parameters are given as follows.
All ground users are placed on the ground within the area of $500 \times 500$ ${\rm{m}^2}$.
Two UAVs are dispatched to provide communications to two ground users with the existence with two potential eavesdroppers.
{\color{black}
Furthermore, we assume that the two UAVs' initial locations and final locations in 2D area are $\mathbf{q}_{1}^0 = (0,0)$, $\mathbf{q}_{1}^F = (500,0)$, $\mathbf{q}_{2}^0 = (0,500)$, and $\mathbf{q}_{2}^F = (500,500)$, respectively. }
The communication bandwidth is 2 MHz with a carrier center frequency at 2 GHz, the number of subcarrier $N_\mathrm{F} = 16$, and the noise power on each subcarrier is $-100$ dBm with channel gain $\beta_0 = -50$ dB at the reference distance $d_0 = 1$ m.
Therefore, the channel gain-to-noise ratio at the reference distance is $\gamma_0 = 80$ $\rm{dB}$ \cite{Zhong_Cooperative_Jamming}.
All UAVs' maximum transmission power are set as $P_\mathrm{peak} = 30$ dBm \cite{Lee_cooperative_jamming}.
Both of the UAV's maximum flying speed is $v_{\max} = 20$ m/s with a fixed altitude of $H = 100$ m, and the safety distance between any two UAVs is $D_\mathrm{S}= 50$ m.
{\color{black}The centres of two NFZs are $\mathbf{w}_{\mathrm{NF}}^1 = (150,325)$ and $\mathbf{w}_{\mathrm{NF}}^2 = (350,325)$, respectively, and the radii of both NFZs are the same with $Q_\mathrm{NF} = 60$ m.}
In all considered scenarios, two users are located on $\mathbf{w}_{k1_U} = (50,50)$ and $\mathbf{w}_{k2_U} = (400,450)$, respectively.
{\color{black}Furthermore, as in \cite{Zhong_Cooperative_Jamming,xiao2018secrecy,a_li_cooperative_2019}, we assume that three eavesdroppers' locations are exactly known as $\mathbf{w}_{1_E} = (70,70)$, $\mathbf{w}_{2_E} = (150,250)$, and $\mathbf{w}_{3_E} = (250,150)$.}
For illustration, all the trajectories are sampled every second.
\vspace*{-1mm}

In our simulation, we compare the performance of the proposed algorithm, denoted as PA with the other two baseline schemes: a) \textit{No jammer (NJ)}, which all UAVs serve as information UAV and do not transmit artificial noise to eavesdroppers \cite{Zhang2018Securing,cui2018robust,Yuanxin_Cai}.
Since the problem for PA subsumes the NJ scheme as a subcase, the average minimum secrecy rate per user in NJ can be achieved by solving subproblem 1 and subproblem 3 with the settings of jamming power as zero.
(b) \textit{Single-purpose UAV (SP)}, which one UAV can provide communication while the other is acting as a jammer at all time \cite{a_li_cooperative_2019,Zhong_Cooperative_Jamming,Lee_cooperative_jamming}.
SP is also a subcase for PA, and the average minimum secrecy rate in PA can be obtained by solving the problem in \eqref{eqn:problem 1} with fixing UAV 1 as a jammer and UAV 2 as a base station.

\vspace*{-3mm}

\subsection{Proposed Trajectories with Different Mission Time Durations}\vspace*{-1.5mm}

\begin{figure}[t]
  \centering

  \begin{minipage}{13cm}
  \centering
  \includegraphics[width=4.2 in]{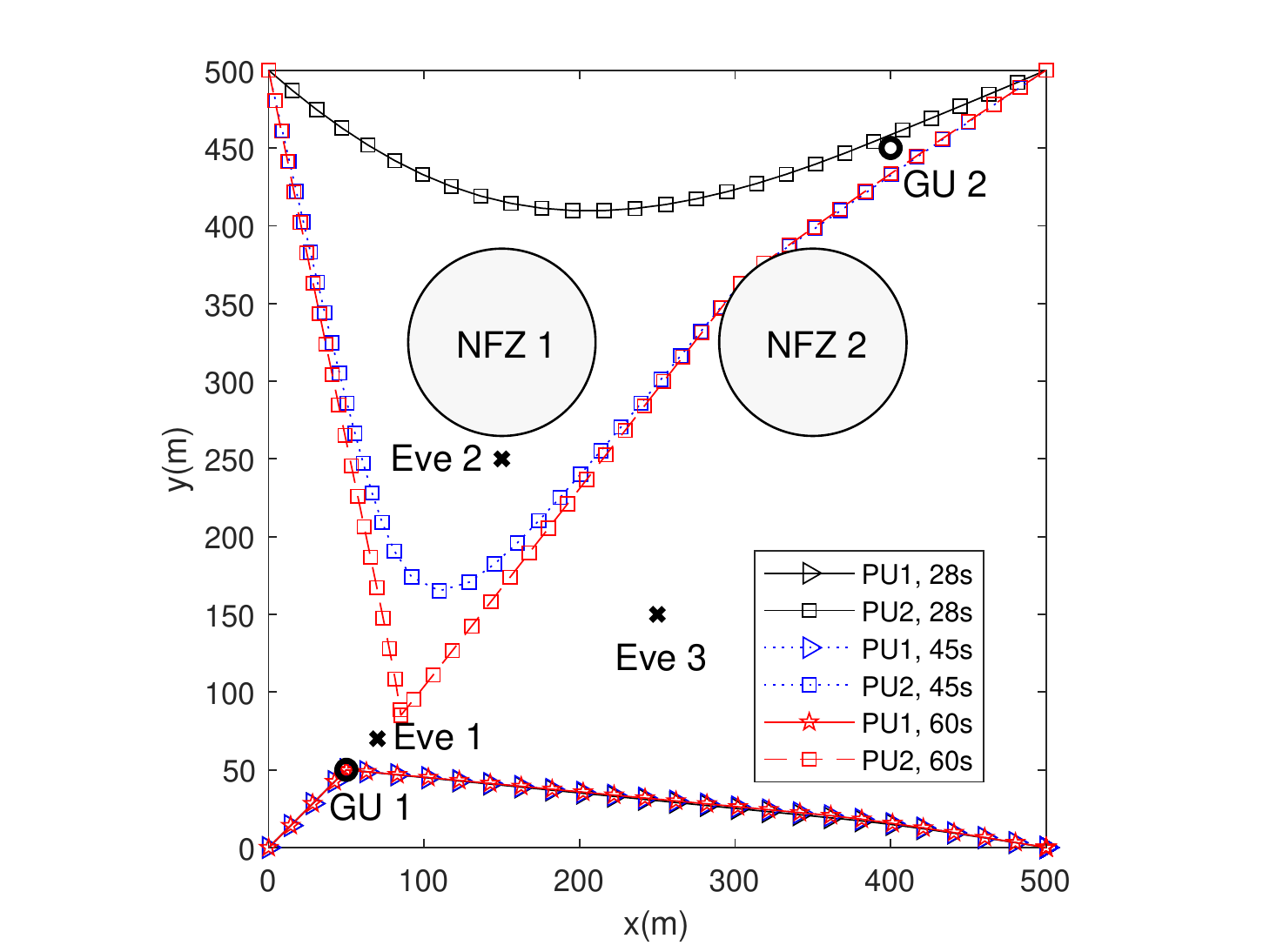}

  \end{minipage}

    \begin{minipage}{13cm}
  \centering
  \includegraphics[width=4.3 in]{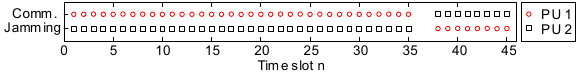}
  \end{minipage}\vspace*{-3mm}
  \caption{{\color{black}The upper half of the figure shows the proposed algorithm trajectories with different mission time durations. The lower half of the figure depicts the roles of the two UAVs for the case of $T=45$ s.}}\vspace*{-9mm}\label{fig:different time}

\end{figure}

{\color{black}
Fig. \ref{fig:different time} illustrates UAVs' trajectories of our proposed scheme with three different mission time durations, $T = 28$ $\rm{s}$, $T = 45$ $\rm{s}$, and $T = 60$ $\rm{s}$, respectively.
The proposed trajectories in this figure for UAV 1 and UAV 2 are denoted as PU 1 and PU 2, respectively.
It is observed that when $T$ is relatively small (e.g. $T = 28$ $\rm{s}$), UAV 1 first goes directly towards ground user 1 (GU 1) and hovers over it before flying back straightly to the destination, whereas the behavior of UAV 2 is totally different.
This is because all three eavesdroppers are closer to GU 1 compared to GU 2.
Specifically, eavesdropper 1 is in close proximity to GU 1 which incurs a high risk in information leakage.
Therefore, UAV 2 first acts as a jammer and flies close to all the eavesdroppers as possible at the beginning to help UAV 1 for secure communications.
After UAV 1 conveys enough secure data to GU 1, UAV 1 switches its role to a jamming UAV while on its way back to its final location.
Concurrently, to achieve fairness in resource allocation, UAV 2 flies towards GU 2 and transmits information to GU 2 when they are close enough to each other.
In addition, both UAVs fly with the maximum speed $v_{\max} = 20$ m/s to establish a shorter LoS link to the user/eavesdropper as fast as possible.
Moreover, we can observe that with a sufficiently long time duration (e.g. $T = 45$ $\rm{s}$), the behaviors of the UAVs alter to fully exploit the degrees of freedom brought by the additional time.
For the ease of illustration, the lower half of Fig. \ref{fig:different time} shows the roles of UAVs across time for the case of $T = 45$ $\rm{s}$.
In particular, for $n\le35$, UAV 2 first cruises towards eavesdropper 1 and serves as a jammer to assist the secure communications between UAV 1 and GU 1.
Besides, with the help of UAV 2, UAV 1 also takes the shortest path cruising towards GU 1 and hovers over it for efficient communication.
From $35<n<38$, both UAVs prepare role switching by navigating themselves to their desired positions.
During this short period of time, the channels between UAVs and any of the eavesdroppers are better than that of all users even if jamming is performed.
To prevent information leakage, both UAVs would not communicate to any ground user, as revealed in Lemma 3.
Therefore, there is neither communication nor jamming in the system in this period.
Furthermore, when UAV 2 is close enough to GU 2 for efficient secure communication at $n \ge 38$, UAV 1 starts serving as a jammer on its journey to final destination to protect GU 2 against eavesdropping.
It is also observed that when the straight direction between a UAV and the desired location is blocked by a NFZ, the UAV's trajectory of the proposed scheme would take the shortest path along the tangential line of the NFZ.
Also, the roles of both the UAVs remain the same before cruising back to their corresponding destinations.
We can also observe that when the mission time duration is sufficiently long (e.g. $T = 60$ $\rm{s}$), UAV 2 would spend more time and hover over at some locations close to eavesdropper 1 to fully exploit the additional time as a jammer for efficient jamming.
Moreover, UAV 2 keeps safety distance with UAV 1 and there is no collision between them thanks to the collision avoidance constraint $\mathrm{C9}$ in the problem formulation.}
\vspace*{-4mm}
\subsection{Baseline Trajectories with Fixed Mission Duration Time}\vspace*{-1.5mm}

\vspace*{0mm}
\begin{figure}[t]
  \centering
  \includegraphics[width=4.2 in]{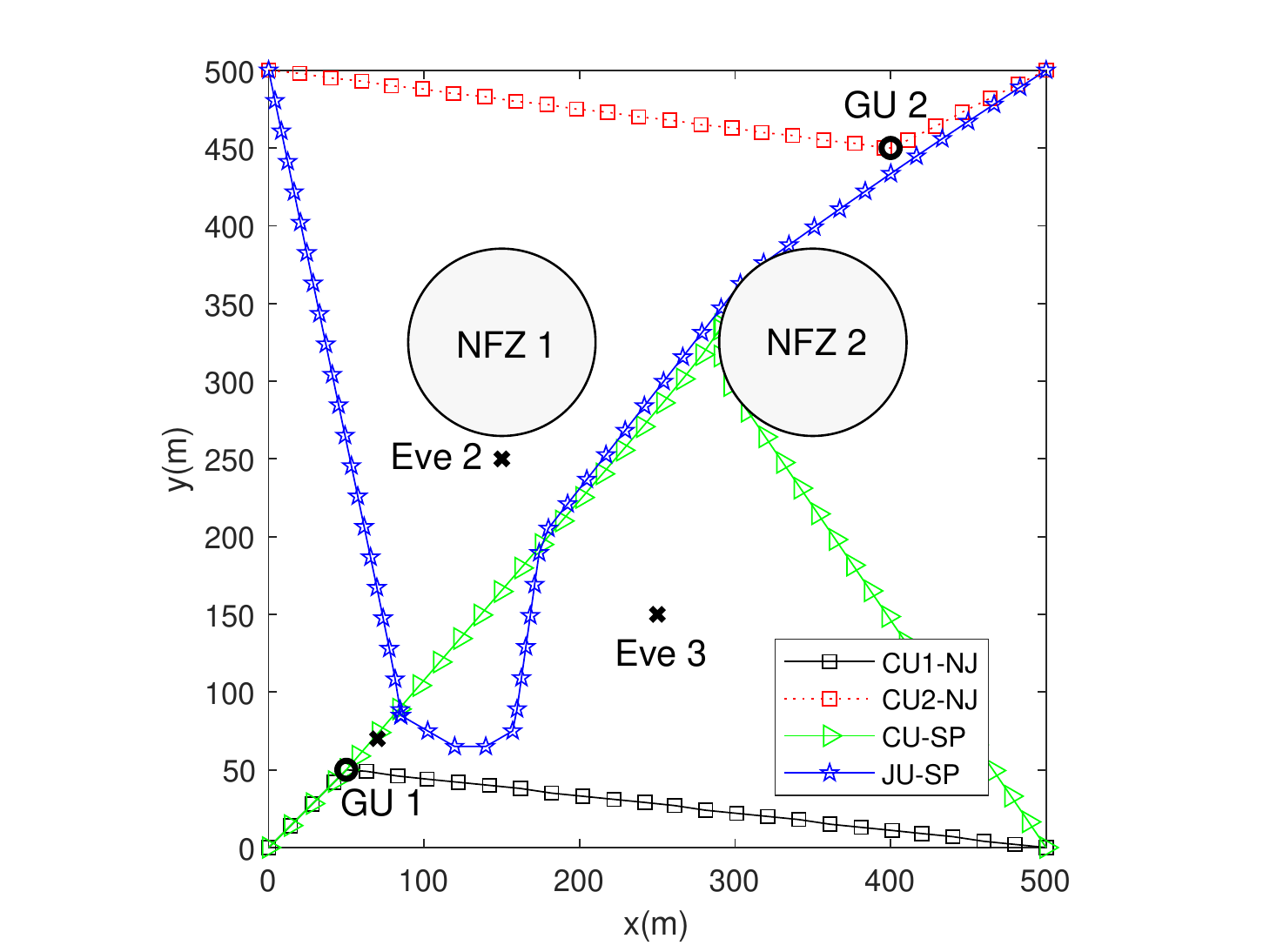}\vspace*{-6mm}
  \caption{{\color{black}Trajectories of single-purpose UAVs for different schemes.}}\vspace*{-6mm}
  \label{fig:Single function}\vspace*{-4mm}
\end{figure}

{\color{black}
Fig. \ref{fig:Single function} shows the trajectories of a single-purpose UAV system for a mission duration time of $T=60$ $\rm{s}$.
For baseline 1 (NJ), the trajectories of communication UAV 1 and communication UAV 2 are denoted as CU1-NJ and CU2-NJ, respectively.
For baseline 2 (SP), the trajectories of communication UAV and jammer UAV are denoted as CU-SP and JU-SP, respectively.
It can be observed that for baseline 1, UAV 1 and UAV 2 fly directly to GU 1 and GU 2 with the maximum speed $v_{\max} = 20$ m/s, respectively.
Besides, each UAV hovers over their corresponding desired GU as long as possible within the allowed mission duration time.
Then, both of the UAVs fly directly to the final location.
Different from baseline 1, the single purpose communication UAV CU-SP in baseline 2 just hovers over GU 1 for a short period duration time and then flies directly to GU 2 from the left hand side of NFZ 2 for establishing efficient communication between them.
This is because CU-SP needs to fly towards and communicate to a far away user, i.e., GU 2, even if another single-purpose jamming UAV JU-SP is closer to this remote user.
In this scheme, the only communication UAV has no choice but fly closer to the desired user to establish a strong LoS link to provide secure communication.
At the same time, the JU-SP UAV flies close to eavesdropper 1 for jamming at the beginning to relieve the system bottleneck created by eavesdropper 1.
Then, the JU-SP UAV flies close to the centroid location formed by the three eavesdroppers to improve secrecy rate communication.
In particular, to avoid UAVs collision, JU-SP moves away from its hovering location to give way for CU-SP.
This is because in baseline 2, there is only one communication UAV for serving all users which imposes a very stringent restriction in utilizing the system resources for maximizing the average minimum secrecy rate.
}

\vspace*{-6mm}
\subsection{Average Minimum Secrecy Rate versus Mission Duration Time}\vspace*{-1.5mm}

\vspace*{0mm}
\begin{figure}[t]
  \centering
  \includegraphics[width=3.9 in]{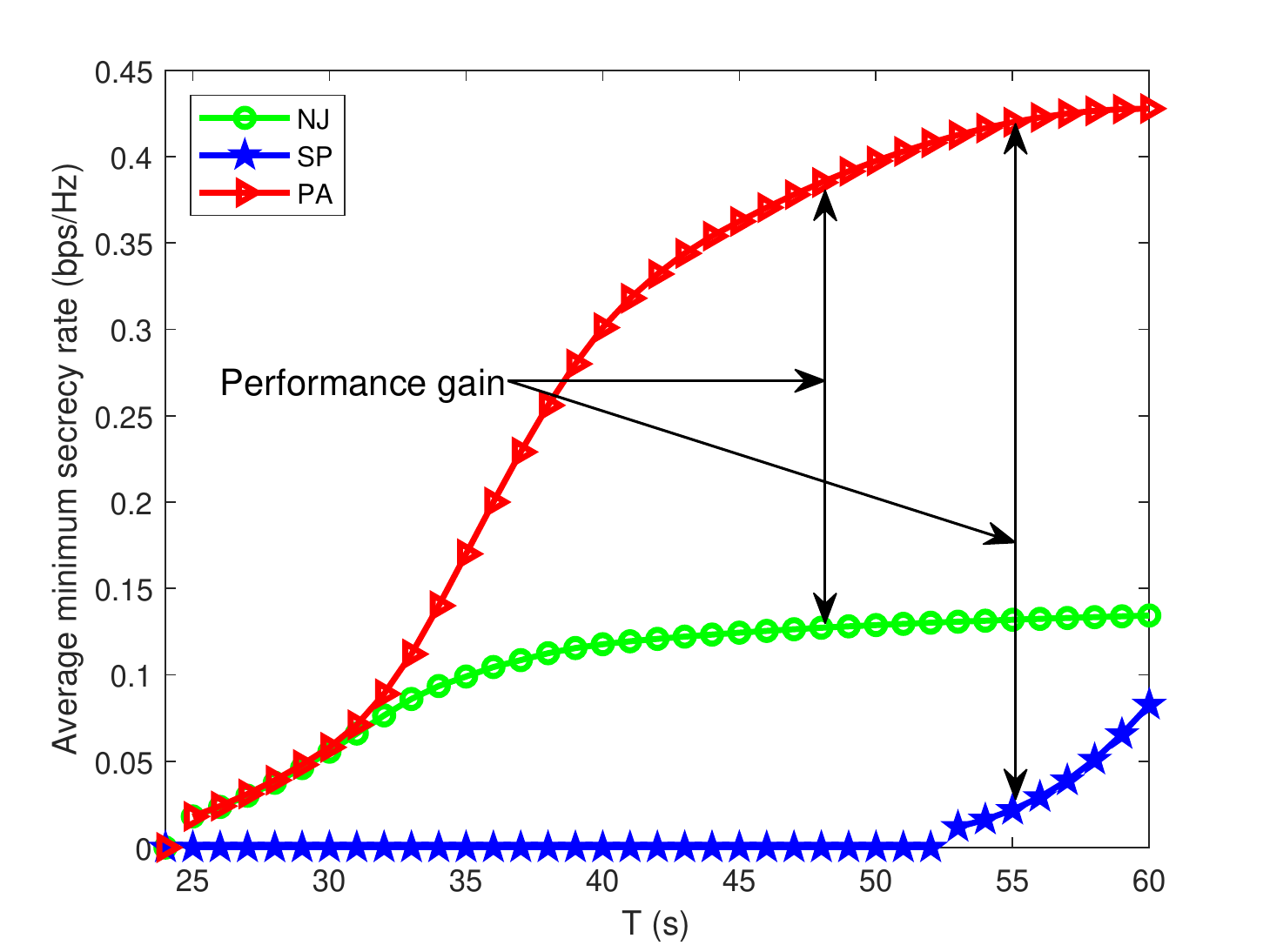}\vspace*{-8mm}
  \caption{{\color{black}Average minimum secrecy rate versus the mission time durations with different schemes.}}\vspace*{-10mm}
  \label{fig:MAS_rate_diff_time}
\end{figure}

{\color{black}
Fig. \ref{fig:MAS_rate_diff_time} depicts the average minimum secrecy rate versus the mission duration time $T$ for our proposed algorithm and the other two baselines with three eavesdroppers.
It is observed that for all the considered algorithms, the average minimum secrecy rate is virtually zero when the mission time duration is less than 25 $\rm{s}$.
In fact, a small time duration would lead to an infeasible result for all the algorithms, since both the UAVs cannot fly back to their final locations even if there is no NFZ and they cruise with the maximum aviation speed $v_{\max} = 20$ m/s.
Also, it can be observed that the average minimum secrecy rates achieved by PA and NJ both increase with mission duration time $T$.
However, for SP, the average minimum secrecy rate is zero until the mission duration time $T$ is longer than 53 $\rm{s}$.
This is because the only communication UAV in SP has insufficient time to reach closer to GU 2 and to provide secure communication.
As a result, the UAV would not transmit anything to avoid information leakage, as revealed in Lemma 3.
Furthermore, we can observe that the system performance of PA increases sharply from 30 $\rm{s}$ to 40 $\rm{s}$ while the average minimum secrecy rate of NJ scales up slowly.
In fact, a sufficiently long time duration grants a UAV more time to reach and stay close to its desired location to enjoy a short distance communication.
Besides, the proposed role switching for UAVs grants the UAVs much higher flexibility to provide communication or to act as a jammer, which can save considerable time for the UAVs to fly to their desired destinations.
However, the system performance gain brought by the extra time is diminishing for both PA and NJ.
In other words, for a given maximum transmit power, the performance bottleneck created by limited time duration is relieved as each GU can be served with a sufficiently long time while enjoying excellent channel conditions with UAVs.
}

\vspace*{-6mm}
\subsection{Average Minimum Secrecy Rate versus Maximum Transmission Power}\vspace*{-1.5mm}

\vspace*{0mm}
\begin{figure}[t]
  \centering
  \includegraphics[width=3.9 in]{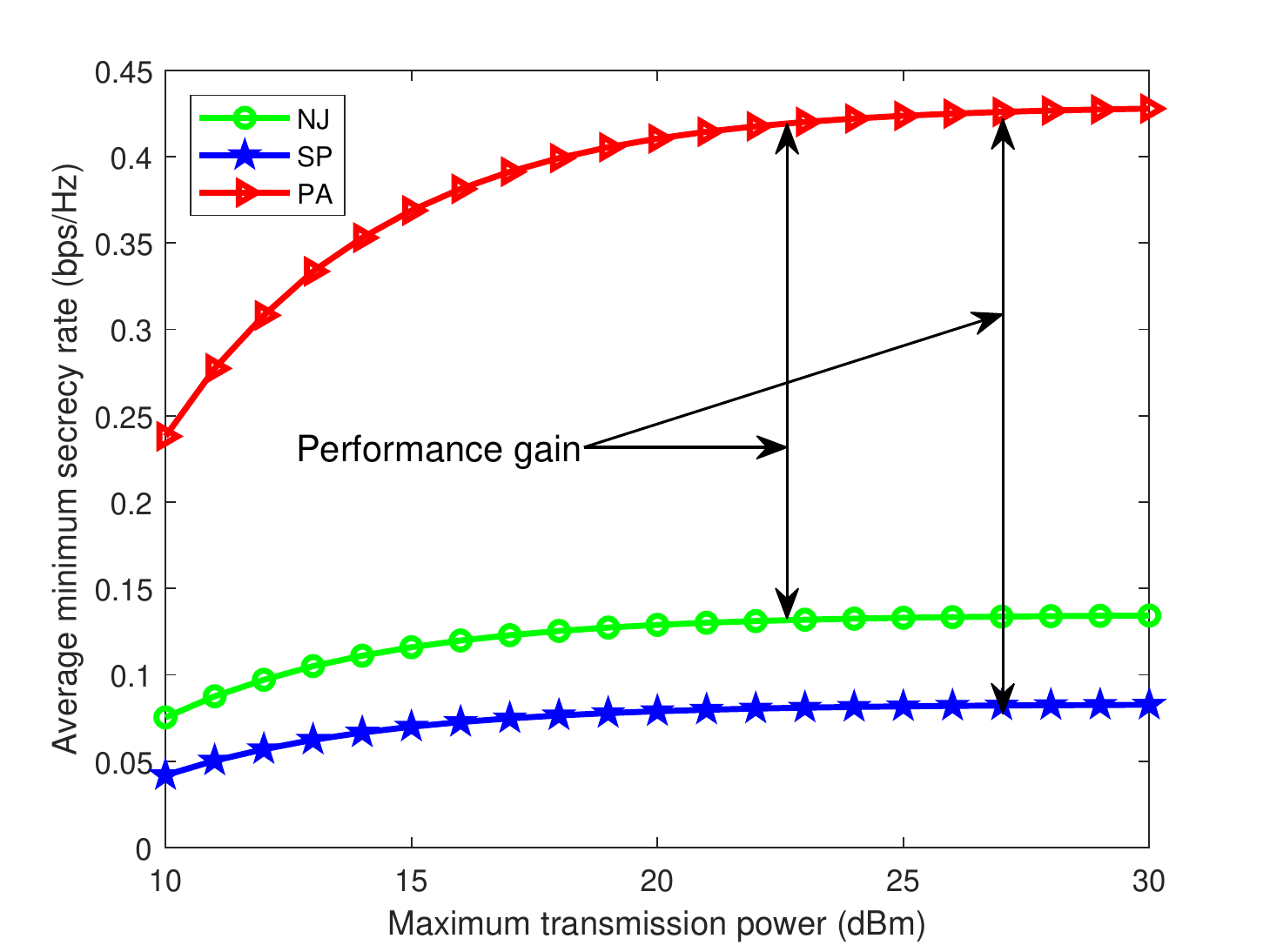}\vspace*{-8mm}
  \caption{{\color{black}Average minimum secrecy rate versus the maximum transmission power with different schemes.}}\vspace*{-10mm}
  \label{fig:MAS_rate_diff_func}
\end{figure}

{\color{black}
Fig. \ref{fig:MAS_rate_diff_func} demonstrates the achieved average minimum secrecy rate versus the maximum transmission power for all schemes for $T=60$ $\rm{s}$.
It can be observed that the average minimum secrecy rates achieved by all schemes increase with the maximum transmission power at the beginning.
Indeed, a higher data rate can be achieved with a higher power when the UAVs are close to their desired users.
However, the system performance of all three algorithms is saturated when the maximum transmission power is sufficiently high.
This is because eavesdropper 1 is in the neighbourhood of GU 1, a higher transmission power may incur a higher information leakage and the UAV would clip its transmit power at a certain level despite there is still transmit power available.
Moreover, it is interesting to see that the performance gain achieved by our proposed algorithm PA are quite large at all times compared to other two baselines which unveils that role switching among UAVs is the key for improving the system performance.
In addition, the average minimum secrecy rate of both SP and NJ schemes scale slowly w.r.t. the available power compared to the proposed scheme as they are less efficient in exploiting the extra transmission power for improving the system performance.
Also, despite a higher maximum transmission power grants all UAVs higher capabilities to improve secure communication system performance, single-purpose UAVs perform the worst among all the considered schemes as they are not competent for time-critical mission in secure communication systems due to the distributed nature of GUs.}

\vspace*{-5mm}
\section{Conclusion}\vspace*{-2mm}

In this paper, we presented a new communication approach with multi-purpose UAVs which can either provide communication to users or serve as jammers to transmit noise signal to eavesdroppers.
By exploiting the UAVs' high mobility, we maximize the average minimum secrecy rate by jointly optimizing the communication user scheduling, communication power allocation, jamming policy, and the trajectories of UAVs, while taking into account the safety distance between any two UAVs, restricted flight in NFZs, the maximum UAV cruising speed, and initial/final UAV locations.
A suboptimal solution of resource allocation for secure multi-UAV communication systems was derived by utilizing dual decomposition and SCA with a polynomial time computational complexity.
Numerical results demonstrated that our proposed design of multi-purpose UAV communication system can significantly increase the minimum secrecy data rate compared to various baseline schemes enabled by dynamically switching the roles of UAVs between communication and jamming.

\vspace*{-5mm}
\section*{Appendix}
\vspace*{-2mm}
\begin{appendices}

\vspace*{-3mm}
\subsection{Proof of Lemma \ref{lemma1}}\label{Proof of Lemma 1}\vspace*{-1.5mm}
\vspace*{-1.5mm}
Denote $\eta_1^*$ and $\eta_2^*$ as the optimal values of problems \eqref{eqn:problem 1} and \eqref{eqn:no nonsm problem 1}, respectively.
Since $[x]^+ \ge x, \forall x$, we have $\eta_1^* \ge \eta_2^*$.

Next, we prove that $\eta_2^* \ge \eta_1^*$ holds either.
Denote $({\mathcal{Q}^*, \mathcal{S}_U^*,\mathcal{S}_J^*, \mathcal{P}_U^*, \mathcal{P}_J^*})$ as the optimal solution to \eqref{eqn:problem 1}, where $\mathcal{P}_U^*  = \{p_{m,k_U,i}^*[n] , \forall n,m,k_U,i\}$.
Define $f(p_{m,k_U,i}[n]) = R_{m,k_U,i}[n] - R'_{m,k_U,k_E,i}[n]$.
We construct a feasible solution $(\hat{\mathcal{Q}}, \hat{\mathcal{S}}_U, \hat{\mathcal{S}}_J, \hat{\mathcal{P}_U}, \hat{\mathcal{P}}_J)$
to \eqref{eqn:no nonsm problem 1}, such that $\hat{\mathcal{Q}} = \mathcal{Q}^*$, $\hat{\mathcal{S}}_U = \mathcal{S}_U^*$, $\hat{\mathcal{S}}_J = \mathcal{S}_J^*$, $\hat{\mathcal{P}}_J = \mathcal{P}_J^*$,
and the elements of $\hat{\mathcal{P}_U}$ can be obtained as: if $f(p_{m,k_U,i}[n]) \ge 0$, $\hat{P}_{m,k_U,i}[n] = P^*_{m,k_U,i}[n]$; otherwise $\hat{P}_{m,k_U,i}[n] = 0$.
Denote $\hat{\eta}$ as the objective value of \eqref{eqn:no nonsm problem 1} attained at $\hat{\mathcal{Q}}$, $\hat{\mathcal{S}}_U$, $\hat{\mathcal{S}}_J$, $\hat{\mathcal{P}_U}$, $\hat{\mathcal{P}_J}$.
Therefore, the newly constructed solution $(\hat{\mathcal{Q}},\hat{\mathcal{S}}_U,\hat{\mathcal{S}}_J, \hat{\mathcal{P}_U}, \hat{\mathcal{P}}_J)$ ensures $\hat{\eta} = \eta_1^*$, but also is feasible to \eqref{eqn:no nonsm problem 1},
which follows that $\eta_2^* \ge \hat{\eta} $ and thus $\eta_2^*  \ge \eta_1^* $. Therefore, $\eta_2^* = \eta_1^* $, which completes the proof.

\vspace*{-6mm}
\subsection{Proof of Lemma \ref{lemma2}}\label{Proof of Lemma 2}
\vspace*{-2.5mm}

The Hessian of $\psi(x, y) \triangleq x\log_2(1 + \frac{\kappa_1 y}{x}) - x \log_2(1 + \frac{\kappa_2 y}{x})$ is given by\vspace*{-2mm}
\begin{equation}
\bigtriangledown^2 \psi(x, y)= \left[
\begin{array}{cc}
\frac{\kappa_2^2y^2}{x^3 {( \frac{\kappa_2y}{x} + 1)^2} }-\frac{\kappa_1^2y^2}{x^3 {( \frac{\kappa_1y}{x} + 1)^2} } & \frac{\kappa_1^2y}{x^2{( \frac{\kappa_1y}{x} + 1)^2}}-\frac{\kappa_2^2y}{x^2 {( \frac{\kappa_2y}{x} + 1)^2}}        \\
\frac{\kappa_1^2y}{x^2{( \frac{\kappa_1y}{x} + 1)^2}} - \frac{\kappa_2^2y}{x^2 {( \frac{\kappa_2y}{x} + 1)^2}}         &  \frac{\kappa_2^2}{x {( \frac{\kappa_2y}{x} + 1)^2}} - \frac{\kappa_1^2}{x {( \frac{\kappa_1y}{x} + 1)^2}}
\end{array}\right].
\end{equation}

{\color{black}
For any $\bold{t}  = [ t_1 , t_2 ]^T$,  we have\vspace*{-2mm}
%
\begin{align}
\bold{t}^T \bigtriangledown^2  \psi(x, y) \bold{t}
=& -\frac{1}{x}\left( \frac{t_1y^2}{x} - t_2 \right)^2
  \left( \frac{\kappa_1^2}{\left( \frac{\kappa_1y}{x} +1 \right)^2} - \frac{\kappa_2^2}{\left( \frac{\kappa_2y}{x} +1 \right)^2} \right).
\end{align}
It is easy to verify that, for $\kappa_1 > \kappa_2 \ge 0$, $x\ge0$, and $y \ge 0$,\vspace*{-1.5mm}
\begin{align}
\frac{\kappa_1^2}{\left( \frac{\kappa_1y}{x} +1 \right)^2} - \frac{\kappa_2^2}{\left( \frac{\kappa_2y}{x} +1 \right)^2}
\ge 0,
\end{align}
so that\vspace*{-5mm}
\begin{align}
\bold{t}^T \bigtriangledown^2  \psi(x, y) \bold{t} \le 0,
\end{align}
when $\kappa_1 > \kappa_2 \ge 0$, $x\ge0$, and $y \ge 0$.
}
Therefore, the Hessian $\psi(x, y)$ is a negative semi-definite matric and $\psi(x, y)$ is a concave function w.r.t. $x$ and $y$.

\vspace*{-7mm}
\subsection{Proof of Lemma \ref{lemma3}}\label{Proof of Lemma 3}
\vspace*{-2.5mm}
If UAV $m$ allocates a non-negative power for user $k_U$ on subcarrier $i$ in time slot $n$, i.e., $s_{m,k_U,i}[n] = 1$ and $\widetilde{p}_{m,k_U,i}[n] > 0$, under the condition  $\mathcal{H}_{m,k_U,i}[n] \le \mathcal{H}'_{m,k_E,i}[n]$, we have $\bigg [ \tilde{R}_{m,k_U,i}[n] - \tilde{R}'_{m,k_U,k_E,i}[n] \bigg ] \le 0$, which leads to a smaller objective value in \eqref{eqn:problem 1 tilde p}.
Besides, the solution set becomes smaller as some power is wasted without improving the objective value.
Hence, $\widetilde{p}_{m,k_U,i}[n] > 0$ for  $\mathcal{H}_{m,k_U,i}[n] < \mathcal{H}'_{m,k_E,i}[n]$ is not an optimal solution.

\end{appendices}


\begin{thebibliography}{10}
\providecommand{\url}[1]{#1}
\csname url@samestyle\endcsname
\providecommand{\newblock}{\relax}
\providecommand{\bibinfo}[2]{#2}
\providecommand{\BIBentrySTDinterwordspacing}{\spaceskip=0pt\relax}
\providecommand{\BIBentryALTinterwordstretchfactor}{4}
\providecommand{\BIBentryALTinterwordspacing}{\spaceskip=\fontdimen2\font plus
\BIBentryALTinterwordstretchfactor\fontdimen3\font minus
  \fontdimen4\font\relax}
\providecommand{\BIBforeignlanguage}[2]{{%
\expandafter\ifx\csname l@#1\endcsname\relax
\typeout{** WARNING: IEEEtran.bst: No hyphenation pattern has been}%
\typeout{** loaded for the language `#1'. Using the pattern for}%
\typeout{** the default language instead.}%
\else
\language=\csname l@#1\endcsname
\fi
#2}}
\providecommand{\BIBdecl}{\relax}
\BIBdecl

\bibitem{li}
R.~{Li}, Z.~{Wei}, L.~{Yang}, D.~W. {Kwan Ng}, N.~{Yang}, J.~{Yuan}, and
  J.~{An}, ``Joint trajectory and resource allocation design for {UAV}
  communication systems,'' in \emph{2018 IEEE Globecom Workshops (GC Wkshps)},
  Dec. 2018, pp. 1--6.

\bibitem{j_an}
J.~{An}, K.~{Yang}, J.~{Wu}, N.~{Ye}, S.~{Guo}, and Z.~{Liao}, ``Achieving
  sustainable ultra-dense heterogeneous networks for {5G},'' \emph{IEEE Commun.
  Mag.}, vol.~55, no.~12, pp. 84--90, Dec. 2017.

\bibitem{K_Yang}
K.~{Yang}, N.~{Yang}, N.~{Ye}, M.~{Jia}, Z.~{Gao}, and R.~{Fan},
  ``Non-orthogonal multiple access: Achieving sustainable future radio
  access,'' \emph{IEEE Commun. Mag.}, vol.~57, no.~2, pp. 116--121, Feb. 2019.

\bibitem{X_Gao}
X.~{Gao}, P.~{Wang}, D.~{Niyato}, K.~{Yang}, and J.~{An}, ``Auction-based time
  scheduling for backscatter-aided {RF}-powered cognitive radio networks,''
  \emph{IEEE Trans. Wireless Commun.}, vol.~18, no.~3, pp. 1684--1697, Mar.
  2019.

\bibitem{wong_2017_Key_Technologies_for_5G}
V.~W. Wong and L.-C. Wang, \emph{Key Technologies for {5G} Wireless
  Systems}.\hskip 1em plus 0.5em minus 0.4em\relax Cambridge University Press,
  2017.

\bibitem{Zeng_2016_Unmanned_Aerial_Vehicles_Magazine}
Y.~Zeng, R.~Zhang, and T.~J. Lim, ``Wireless communications with unmanned
  aerial vehicles: opportunities and challenges,'' \emph{IEEE Commun. Mag.},
  vol.~54, no.~5, pp. 36--42, May 2016.

\bibitem{Zeng_2016_Throughput_Maximization_for_UAV}
------, ``Throughput maximization for {UAV}-enabled mobile relaying systems,''
  \emph{IEEE Trans. Commun.}, vol.~64, no.~12, pp. 4983--4996, Dec. 2016.

\bibitem{xiao2018secrecy}
L.~{Xiao}, Y.~{Xu}, D.~{Yang}, and Y.~{Zeng}, ``Secrecy energy efficiency
  maximization for {UAV}-enabled mobile relaying,'' \emph{IEEE Transactions on
  Green Communications and Networking}, pp. 1--1, Oct. 2019.

\bibitem{Device_to_Device_Communications_Mozaffari}
M.~Mozaffari, W.~Saad, M.~Bennis, and M.~Debbah, ``Unmanned aerial vehicle with
  underlaid device-to-device communications: Performance and tradeoffs,''
  \emph{IEEE Trans. Wireless Commun.}, vol.~15, no.~6, pp. 3949--3963, Jun.
  2016.

\bibitem{Robust_sun_2019}
X.~{Sun}, C.~{Shen}, D.~W.~K. {Ng}, and Z.~{Zhong}, ``Robust trajectory and
  resource allocation design for secure {UAV}-aided communications,'' in
  \emph{2019 IEEE Intern. Conf. on Commun. Workshops (ICC Workshops)}, May
  2019, pp. 1--6.

\bibitem{Zeng_2017_Completion_Time_Minimization_in_UAV}
Y.~Zeng, X.~Xu, and R.~Zhang, ``Trajectory design for completion time
  minimization in {UAV}-enabled multicasting,'' \emph{IEEE Trans. Wireless
  Commun.}, vol.~17, no.~4, pp. 2233--2246, Apr. 2018.

\bibitem{Sun_2019_Physical}
X.~{Sun}, D.~W.~K. {Ng}, Z.~{Ding}, Y.~{Xu}, and Z.~{Zhong}, ``Physical layer
  security in {UAV} systems: Challenges and opportunities,'' \emph{IEEE
  Wireless Commun. Mag.}, vol.~26, no.~5, pp. 40--47, Oct. 2019.

\bibitem{wu2019safeguarding}
Q.~Wu, W.~Mei, and R.~Zhang, ``Safeguarding wireless network with {UAV}s: A
  physical layer security perspective,'' \emph{arXiv preprint
  arXiv:1902.02472}, 2019.

\bibitem{Zhang2018Securing}
G.~{Zhang}, Q.~{Wu}, M.~{Cui}, and R.~{Zhang}, ``Securing {UAV} communications
  via joint trajectory and power control,'' \emph{IEEE Trans. Wireless
  Commun.}, vol.~18, no.~2, pp. 1376--1389, Feb. 2019.

\bibitem{cui2018robust}
M.~{Cui}, G.~{Zhang}, Q.~{Wu}, and D.~W.~K. {Ng}, ``Robust trajectory and
  transmit power design for secure {UAV} communications,'' \emph{IEEE Trans.
  Veh. Technol.}, vol.~67, no.~9, pp. 9042--9046, Sep. 2018.

\bibitem{Yuanxin_Cai}
Y.~{Cai}, Z.~{Wei}, R.~{Li}, D.~W. {Kwan Ng}, and J.~{Yuan}, ``Energy-efficient
  resource allocation for secure {UAV} communication systems,'' in \emph{2019
  IEEE Wireless Commun. and Networking Conf. (WCNC)}, Apr. 2019, pp. 1--8.

\bibitem{a_li_cooperative_2019}
A.~{Li}, Q.~{Wu}, and R.~{Zhang}, ``{UAV}-enabled cooperative jamming for
  improving secrecy of ground wiretap channel,'' \emph{IEEE Wireless Commun.
  Lett.}, vol.~8, no.~1, pp. 181--184, Feb. 2019.

\bibitem{Zhong_Cooperative_Jamming}
C.~{Zhong}, J.~{Yao}, and J.~{Xu}, ``Secure {UAV} communication with
  cooperative jamming and trajectory control,'' \emph{IEEE Commun. Lett.},
  vol.~23, no.~2, pp. 286--289, Feb. 2019.

\bibitem{Y_Li_Access}
Y.~{Li}, R.~{Zhang}, J.~{Zhang}, S.~{Gao}, and L.~{Yang}, ``Cooperative jamming
  for secure {UAV} communications with partial eavesdropper information,''
  \emph{IEEE Access}, pp. 1--1, Jul. 2019.

\bibitem{Lee_cooperative_jamming}
H.~{Lee}, S.~{Eom}, J.~{Park}, and I.~{Lee}, ``{UAV}-aided secure
  communications with cooperative jamming,'' \emph{IEEE Trans. Veh. Technol.},
  vol.~67, no.~10, pp. 9385--9392, Oct. 2018.

\bibitem{Zhang_2017_Securing_UAV_Communications}
G.~{Zhang}, Q.~{Wu}, M.~{Cui}, and R.~{Zhang}, ``Securing {UAV} communications
  via trajectory optimization,'' in \emph{GLOBECOM 2017 - 2017 IEEE Global
  Communications Conference}, Dec. 2017, pp. 1--6.

\bibitem{Valavanis:2014:HUA:2692452}
K.~P. Valavanis and G.~J. Vachtsevanos, \emph{Handbook of Unmanned Aerial
  Vehicles}.\hskip 1em plus 0.5em minus 0.4em\relax Springer Publishing
  Company, Incorporated, 2014.

\bibitem{zhao_chen_yu_2017}
P.~Zhao, W.~Chen, and W.~Yu, ``Guidance law for intercepting target with
  multiple no-fly zone constraints,'' \emph{The Aeronautical Journal}, vol.
  121, no. 1244, pp. 1479--1501, Aug. 2017.

\bibitem{Gao_Access_NFZ}
Y.~{Gao}, H.~{Tang}, B.~{Li}, and X.~{Yuan}, ``Joint trajectory and power
  design for {UAV}-enabled secure communications with no-fly zone
  constraints,'' \emph{IEEE Access}, vol.~7, pp. 44\,459--44\,470, Apr. 2019.

\bibitem{Multiuser_MISO_NFZ}
D.~Xu, Y.~Sun, D.~W.~K. Ng, and R.~Schober, ``Multiuser {MISO} {UAV}
  communications in uncertain environments with no-fly zones: Robust trajectory
  and resource allocation design,'' \emph{arXiv preprint arXiv:1905.10731v2},
  2019.

\bibitem{Wu_OFDM_UAV_2017}
Q.~{Wu} and R.~{Zhang}, ``Common throughput maximization in {UAV}-enabled
  {OFDMA} systems with delay consideration,'' \emph{IEEE Trans. Commun.},
  vol.~66, no.~12, pp. 6614--6627, Dec. 2018.

\bibitem{Guidance_law_circular_NFZ}
W.~Yu and W.~Chen, ``Guidance law with circular no-fly zone constraint,''
  \emph{Nonlinear Dynamics}, Nov. 2014.

\bibitem{Small_UAV_NFZ}
G.~{Ducard}, K.~C. {Kulling}, and H.~P. {Geering}, ``Evaluation of reduction in
  the performance of a small {UAV} after an aileron failure for an adaptive
  guidance system,'' in \emph{2007 American Control Conference}, Jul. 2007, pp.
  1793--1798.

\bibitem{NFZ_Liang}
Z.~Liang and Z.~Ren, ``Tentacle-based guidance for entry flight with no-fly
  zone constraint,'' \emph{Journal of Guidance, Control, and Dynamics},
  vol.~41, no.~4, pp. 996--1005, Dec. 2018.

\bibitem{Colpaert_2018}
A.~Colpaert, E.~Vinogradov, and S.~Pollin, ``Aerial coverage analysis of
  cellular systems at {LTE} and mmwave frequencies using {3D} city models,''
  \emph{Sensors}, vol.~18, no.~12, p. 4311, Dec. 2018.

\bibitem{Hourani_LAP}
A.~{Al-Hourani}, S.~{Kandeepan}, and S.~{Lardner}, ``Optimal {LAP} altitude for
  maximum coverage,'' \emph{IEEE Wireless Commun. Lett.}, vol.~3, no.~6, pp.
  569--572, Dec. 2014.

\bibitem{Zeng_2017_Energy_Efficient_UAV}
Y.~Zeng and R.~Zhang, ``Energy-efficient {UAV} communication with trajectory
  optimization,'' \emph{IEEE Trans. Wireless Commun.}, vol.~16, no.~6, pp.
  3747--3760, Jun. 2017.

\bibitem{3d_Solar_2019}
Y.~{Sun}, D.~{Xu}, D.~W.~K. {Ng}, L.~{Dai}, and R.~{Schober}, ``Optimal
  {3D}-trajectory design and resource allocation for solar-powered {UAV}
  communication systems,'' \emph{IEEE Trans. Commun.}, vol.~67, no.~6, pp.
  4281--4298, Jun. 2019.

\bibitem{Propulsion_Energy_Eom}
S.~{Eom}, H.~{Lee}, J.~{Park}, and I.~{Lee}, ``{UAV}-aided wireless
  communication design with propulsion energy constraint,'' in \emph{2018 IEEE
  Intern. Conf. on Commun. (ICC)}, May 2018, pp. 1--6.

\bibitem{Joint_Altitude_Yang}
Z.~Yang, C.~Pan, M.~Shikh-Bahaei, W.~Xu, M.~Chen, M.~Elkashlan, and
  A.~Nallanathan, ``Joint altitude, beamwidth, location, and bandwidth
  optimization for {UAV}-enabled communications,'' \emph{IEEE Commun. Lett.},
  vol.~22, no.~8, pp. 1716--1719, Aug. 2018.

\bibitem{doppler_Lim}
J.~{Lim}, S.~{Kim}, and D.~{Shin}, ``Two-step doppler estimation based on
  intercarrier interference mitigation for {OFDM} radar,'' \emph{IEEE Antennas
  and Wireless Propagation Letters}, vol.~14, pp. 1726--1729, Apr. 2015.

\bibitem{doppler_Jionghui}
J.~Li, Y.~Zhang, Y.~Zhang, W.~Xiong, Y.~Huang, and Z.~Wang, ``Fast tracking
  doppler compensation for {OFDM}-based {LEO} satellite data transmission,'' in
  \emph{2016 2nd IEEE Intern. Conf. on Comput. and Commun. (ICCC)}, Oct. 2016,
  pp. 1814--1817.

\bibitem{doppler_Wang}
H.~Wang and Q.~Zhang, ``\BIBforeignlanguage{Chinese}{The doppler effect of
  aviation communication in {OFDM} system},''
  \emph{\BIBforeignlanguage{Chinese}{Acta Electronica Sinica}}, vol.~31, no.~6,
  pp. 812 -- 15, Jun. 2003.

\bibitem{Cormen_2001}
C.~E. Leiserson, R.~L. Rivest, T.~H. Cormen, and C.~Stein, \emph{Introduction
  to algorithms}.\hskip 1em plus 0.5em minus 0.4em\relax MIT press Cambridge,
  MA, 2001, vol.~6.

\bibitem{nguyen2018introduction}
L.~D. Nguyen, A.~Kortun, and T.~Q. Duong, ``An introduction of real-time
  embedded optimisation programming for {UAV} systems under disaster
  communication.'' \emph{EAI Endorsed Trans. Indust. Netw. \& Intellig. Syst.},
  vol.~5, no.~17, pp. 1--8, Dec. 2018.

\bibitem{Ng_L}
D.~W.~K. Ng, E.~S. Lo, and R.~Schober, ``Energy-efficient resource allocation
  in {OFDMA} systems with large numbers of base station antennas,'' \emph{IEEE
  Trans. Wireless Commun.}, vol.~11, no.~9, pp. 3292--3304, Jul. 2012.

\bibitem{Boyd}
S.~Boyd and L.~Vandenberghe, \emph{Convex optimization}.\hskip 1em plus 0.5em
  minus 0.4em\relax Cambridge University Press, 2004.

\bibitem{sun_yan_tcom_noma}
Y.~{Sun}, D.~W.~K. {Ng}, J.~{Zhu}, and R.~{Schober}, ``Robust and secure
  resource allocation for full-duplex {MISO} multicarrier {NOMA} systems,''
  \emph{IEEE Trans. Commun.}, vol.~66, no.~9, pp. 4119--4137, Sep. 2018.

\bibitem{channel_power_gain_Gopala}
P.~K. Gopala, L.~Lai, and H.~E. Gamal, ``On the secrecy capacity of fading
  channels,'' \emph{IEEE Trans. Inf. Theory}, vol.~54, no.~10, pp. 4687--4698,
  Oct. 2008.

\bibitem{Derrick_Secure_OFDMA}
D.~W.~K. {Ng}, E.~S. {Lo}, and R.~{Schober}, ``Secure resource allocation and
  scheduling for {OFDMA} decode-and-forward relay networks,'' \emph{IEEE Trans.
  Wireless Commun.}, vol.~10, no.~10, pp. 3528--3540, Oct. 2011.

\bibitem{wei2016power}
Z.~Wei, D.~W.~K. Ng, J.~Yuan, and H.~M. Wang, ``Optimal resource allocation for
  power-efficient {MC-NOMA} with imperfect channel state information,''
  \emph{IEEE Trans. Commun.}, vol.~65, no.~9, pp. 3944--3961, Sep. 2017.

\bibitem{Optimal_Joint_Power_and_Subcarrier_Allocation}
Y.~Sun, D.~W.~K. Ng, Z.~Ding, and R.~Schober, ``Optimal joint power and
  subcarrier allocation for full-duplex multicarrier non-orthogonal multiple
  access systems,'' \emph{IEEE Trans. Commun.}, vol.~65, no.~3, pp. 1077--1091,
  Mar. 2017.

\bibitem{huang2019cognitive}
Y.~Huang, W.~Mei, J.~Xu, L.~Qiu, and R.~Zhang, ``Cognitive {UAV} communication
  via joint maneuver and power control,'' \emph{arXiv preprint
  arXiv:1901.02804}, 2019.

\end{thebibliography}
\end{document}